\documentclass[fleqn,a4paper,12pt]{article}
\usepackage{amssymb}
\usepackage{makeidx}
\usepackage{amsmath}
\usepackage{graphicx}
\usepackage{lscape}
\usepackage{a4}
\usepackage{epsfig}
\begin{document}

\thispagestyle{empty}
\newcommand{\al}{\alpha}
\newcommand{\bt}{\beta}
\newcommand{\s}{\sigma}
\newcommand{\lbd}{\lambda}
\newcommand{\vp}{\varphi}
\newcommand{\va}{\varepsilon}
\newcommand{\gm}{\gamma}
\newcommand{\G}{\Gamma}
\newcommand{\p}{\partial}
\newcommand{\om}{\omega}
\newcommand{\be} {\begin{equation}}
\newcommand{\lo}{\left(}
\newcommand{\ro} {\right)}
\newcommand{\ee} {\end{equation}}
\newcommand{\ba} {\begin{array}}
\newcommand{\ea} {\end{array}}
\newcommand{\ds}{\displaystyle}
\begin{center}
{\large\bf Group classification of systems of non-linear
reaction-diffusion equations with general diffusion matrix.
II. Generalized Turing systems}
\end{center}
\vspace{2mm}
\begin{center}
  A. G. Nikitin\\

Institute of Mathematics of Nat.Acad. Sci of Ukraine, 4
Tereshchenkivska str. 01601 Kyiv, Ukraine
\end{center}
\vspace{2mm}
{\bf Abstract}

Group classification of systems of two coupled nonlinear
reaction-diffusion equation with a diagonal diffusion matrix is
carried out. Symmetries of diffusion systems with singular diffusion matrix
and additional first order derivative terms are described.

\vspace{2mm}

\section{Introduction}

Coupled systems of nonlinear reaction-diffusion equations form the
basis of many models of mathematical biology.  These systems are
widely used in mathematical physics, chemistry and also in social
sciences and many other fields. Such reach spectrum of
applications stimulates numerous thorough investigations of
fundamentals of these equations theory.

In the present paper we continue group classification of systems
of reaction-diffusion equations with general diffusion matrix \be
\label{1.1}\ba{l}
\displaystyle u_t-\Delta_m(A^{11} u+A^{12} v)=f^1(u,v),\\
\displaystyle v_ t-\Delta_m(A^{21} u+A^{22} v)=f^2(u,v) \ea \ee
where $u$ and $v$ are function of $t, x_1, x_2,  \ldots , x_m$,
 $\ A^{11}, A^{12}, A^{21}$ and $A^{22}$ are real constants
and $\Delta_m$ is the Laplace operator in $R^m$.

Up to linear transformations of functions $u,\ v$ and $f^1,\ f^2$
it is sufficient to restrict ourselves to such diffusion matrices
(i.e., matrices whose elements are $A^{11},\cdots,A^{22}$) which
are diagonal, triangular, or are sums of the unit and
antisymmetric matrices. In the last case (\ref{1.1}) can be
reduced to a single equation for a complex function (
generalized complex Ginzburg-Landau (CGL) equation) whose group
classification
 was carried out in paper
\cite{nik1}.

In the present paper we classify equations ({\ref{1.1}) with a
diagonal diffusion matrix. Without loss of generality such
equations can be written as
  \be  \label{1.2}\ba{l}
 u_t-\Delta_m u=f^1(u,v),\\
 v_ t-a\Delta_m v=f^2(u,v) \ea \ee where $a$ is a
constant. The related diffusion matrix is
$A=\left(\ba{cc}1&0\\0&a\ea\right)$.

Just equations of type (\ref{1.2}) are the most popular models of
reaction-diffusion systems first introduced by Turing in 1952 \cite{turing}.
It is practically impossible to enumerate
all fields of
applications of such equations. We restrict ourselves to few
examples only.
\begin{itemize}
\item The complex Ginzburg-Landau (CGL) equation
\be W_t-(1+i\beta)\Delta_2 W=W-(1+i\alpha)|W|^2W\label{la}\ee
can be presented as a system (\ref{1.1})
 where $u$ and $v$ are the real and
imaginary part of $W$. In particular case $\beta=0$ this system
 takes the form (\ref{1.2}).

\item  The primitive predator-prey system which can be defined by
\cite{murry}
\be\label{prey}\ba{l}
u_t-Du_{xx}=-uv,
\\ v_t-\lambda Dv_{xx}=uv\ea
\ee
also appears as an particular subject of our analysis.

\item  The $\lambda -\omega $ reaction-diffusion system \cite{Kop}
\begin{equation}\ba{l}
u_t=\Delta_2 u +\lambda (R)u -\omega (R)v,
\\ v_t=\Delta_2
v+\omega (R)u+\lambda (R)v, \ea \label{d0}
\end{equation}
where $R^2=u^2+v^2, $ is widely used in studies of
reaction-diffusion models, in particular, to describe spiral waves
phenomena \cite{green} .

Symmetries of equations (\ref{d0}) were studied in paper
\cite{AR}. We shall add the results \cite{AR} in the following.

\item The Jackiw-Teitelboim model of two-dimension gravity with
the non-relativistic gauge \cite{martina} appears as a particular
((1+1)-dimensional)
case of the following system:
\be\ba{l}u_t-\Delta_mu=2ku-2u^2v=0,\\
v_t+\Delta_mv=2uv^2-2kv=0.\ea\label{d1}\ee
Symmetries of equations
(\ref{d1}) for $m=1$ were investigated in paper \cite{kra}.
In the following
we complete the results obtained in \cite{kra}.

\end{itemize}

Apparently the first attempt of group classification of equations
(\ref{1.2}) was made by Danilov \cite{danil}. But
the results present in \cite{danil} are rather incomplete.

Group classification of equations (\ref{1.2}) with general non-degenerated
diffusion matrix was announced in \cite{nikwil2} and presented in \cite{nikwil1}.
However, the equivalence relations
where not used systematically there to simplify the equations
which resulted in rather cumbersome form of the classification results.
Moreover due to typographical
errors the tables with classification results present in \cite{nikwil1}
are poorly readable
(see \cite{nik1} for additional
comments).

Group classification of systems of heat equations
\begin{equation}\label{kniazeva}\begin{array}{l}
u_t-\left(f(u,v)u_x\right)_x=0,\\
v_t-\left(g(u,v)v_x\right)_x=0\end{array}\end{equation}
 has been performed in paper \cite{kniazeva}. For some classes of
 functions $f(u,v)$ and $g(u,v)$ equations (\ref{kniazeva}) are
 equivalent to (\ref{1.2}) and the related classification results
 can be compared with ones obtained in the previous section. We will do it in Section 8.

Symmetries of systems of reaction-diffusion equations with a diagonal diffusion matrix
(i.e., of systems (\ref{1.2})) where studied in papers \cite{chern1},
\cite{chern2}.
 We will show in the following that
the classification results obtained in \cite{chern1}, \cite{chern2} are
incomplete and
include many equivalent cases treated as non-equivalent ones.

The problem of group  classification of equations (\ref{1.2})
is still relevant and we will present its solution here. In
addition, we classify equations
(\ref{1.2}) with non-invertible diffusion matrix (i.e., equations
(\ref{1.2}) when parameter $a$ is equal to zero) and also the
following equations with first order derivative terms: \be
\label{1.3} \ba{l}
\displaystyle u_t-\Delta_m u=f^1(u, v), \\
v_t-p_\mu  u_{x_\mu}=f^2(u,v) \ea \ee
 where $u_{x_\mu}=\frac{\p u}{\p x_\mu}$, $p_\mu$ are arbitrary constants
 and summation from 1 to $m$ is imposed over the repeated
 index $\mu$. Moreover, without loss of
generality one can set \be\label{1.31} p_1=p_2=\cdots=p_{m-1}=0, \
p_m=p, \ p=0,1.\ee

In the case $p \equiv 0$ equation (\ref{1.3}) reduces to
(\ref{1.2}) with $a=0$. We notice that this equation is used in such
popular models of mathematical biology as the FitzHung-Naguno
\cite{fitz} and Rinzel-Keller \cite{rinzel} ones.
\section{Equivalence transformations }

The problem of group classification of equations
(\ref{1.2}) and (\ref{1.3}) will be solved up to equivalence
transformations. Clear definition of these transformations is one
of the main points of any classification procedure.

We say the equations \be \label{x.1} \ba{l}
\tilde u_t-\Delta_m \tilde u=\tilde f^1(\tilde u,\tilde v),\\
 \tilde v_ t-a\Delta_m \tilde v=\tilde f^2(\tilde u,\tilde v)
\ea \ee
be equivalent to (\ref{1.2}) if there exist an invertible
transformation $ u \to \tilde u=G(t,x,u,v)$, $v\to \tilde
v=\Phi(t,x,u,v)$, $t \to \tilde t=T(t,x,u,v)$, $x \to \tilde
x=X(t,x,u,v)$ and $f^\al \to \tilde f^\al=F^\al(u,t,x,f^1,f^2)$ which
connects (\ref{1.2}) with (\ref{x.1}). In other words the
equivalence transformations should keep the general form of
equation (\ref{1.2}) but can change the concrete realization
of non-linear terms $f^1$ and $f^2$.

The group of equivalence transformations for equation (\ref{1.2})
can be found using the classical Lie approach and treating $f^1$
and $f^2$ as supplemental dependent variables. In addition to the
obvious symmetry transformations
    \be \label{2.2} t \to t'=t+a,
\quad x_\mu \to x'_\mu=R_{\mu \nu} x_\nu +b_\mu \ee
    where $a, b_\mu$ and $R_{\mu \nu}$ are arbitrary parameters satisfying
$R_{\mu \nu} R_{\mu \lbd}=\delta_{\mu\lbd}$, this group includes
the following transformations
    \be \label{x.2} \ba{l}
u_b\to K^{bc}u_c+b_b, \  f^b \to \lbd^2 K^{bc}f^c,\\
t \to \lbd^{-2} t, \quad x_b \to \lbd^{-1}x_b
 \ea \ee
 and
 \be\label{x.} \ba{l} u_b\to \tilde K^{bc}u_c, \quad f^b \to a^{-1}\tilde K^{bc}f^c,\\
t \to a^{-1} t, \quad x_b \to x_b ,\ \  a\neq 0\ea\ee  where indices $b,c$ take values 1 and 2,
$K^{bc}$ and $\tilde K^{bc}$ are elements of invertible constant matrices
$K$ and $\tilde K$ respectively, moreover, $K$ commutes with $A$,
 and $\tilde K$ satisfy the condition
\[\tilde K A(a) {\tilde K}^{-1}=aA(1/a);\]
 $\lbd \not=0$ and $b_a$ are arbitrary
constants, and we use the temporary notations $u=u_1, v=u_2$.

If parameter $a$ is equal to 1 then $K$ is an arbitrary $2\times 2$ invertible matrix,
and equivalence transformations (\ref{x.}) is trivial.
If $a\neq 1$ then $K$ and $\tilde K$ are arbitrary non-degenerated
diagonal and
 anti-diagonal matrices respectively .

Transformation (\ref{x.}) reduce to the change $\displaystyle a
\to 1/a$ in the related matrix $A$, i.e., to scaling the free parameter
$a$. Thus without loss of generality we can restrict ourselves
to the following values of $a$:
\be\label{x.0}\ba{l} 1. \ a=0, \\ 2.\ -1\leq a<0,\ 0<a<1,\\ 3.\ a=1.\ea\ee

It is possible to show that for $a\neq0$ there is no more extended equivalence
relations valid for arbitrary nonlinearities $f^1$ and $f^2$. If $a=0$ there exist
a powerful equivalence relation $u\to u, \ v\to\varphi(v)$ with an
arbitrary function $\varphi(v)$.
However for some particular functions $f^1$ and $f^2$
the invariance group can be
more extended. In addition to transformations (\ref{x.2})  and (\ref{x.}) it
includes symmetry transformations which does not change the form
of equation (\ref{1.2}). Moreover, for some classes of functions
$f^1, f^2$ equation (\ref{1.2}) admits additional equivalence
transformations (AET) which belong neither to symmetry transformations
nor to transformations of kind (\ref{x.2}), (\ref{x.}).

In spite of the fact that we search for AET  {\it after} description
of symmetries of equations (\ref{1.2}) and specification of
functions $f^1, f^2$, for convenience we present the list of the
additional equivalence transformations in the following formulae:
\be\ba{lll}\label{eqv}
\ 1.&&u\to e^{\rho t}u, \ \ v\to e^{\rho t}v,\\
\ 2.&&u\to u+\omega t,\ v\to v,\\
\ 3.&&u\to u,\ v\to v+\rho t,\\
\ 4.&&u\to u+\mu\rho t,\ v\to e^{-{\rho}t}v,\\
\ 5.&&u\to e^{\rho t}u,\ v\to v-\kappa\rho t, \\
\ 6.&&u\to u,\ v\to v+\rho tu,\\
\ 7.&&u\to e^{2\omega t}u,\ v\to v+\omega {t^2},\\
\ 8.&&u\to  u+\om t^2,\ v\to ve^{2\omega t},\\
\ 9.&&u\to u,\ v\to v-2\rho tu+\rho\delta {t^2},\\
 10.&&u\to e^{2\rho t}u,\ v\to e^{2\rho t}\lo v+
 \omega t u+\rho t^2u\ro,\\
11.&&u\to u+\eta\rho t,\ v\to v-\rho t,\\
12.&&u\to e^{\kappa t}u,\
v\to e^{\kappa t}(v-\nu\kappa t u),\\
13.&&u\to u+2\rho t,\ v\to v+2\rho t
u+2\rho^2{t^2},\\
14.&&u\to e^{\omega t}u, \ \ v\to e^{\rho t}v,\
\\
15.&&u\to e^{\nu\omega t}\lo u\cos(\omega\s t)+v\sin(\omega\s t)\ro,\\
 &&v\to e^{\nu\omega t}\lo v\cos(\omega\s t)-u\sin(\omega\s t)\ro,\\
 16.&&u\to e^{2\omega t}\lo u\cos(\s\omega t^2)-v\sin(\s\omega t^2)\ro,\\
 &&v\to e^{2\omega t}\lo v\cos(\s\omega t^2)+u\sin(\s\omega t^2)\ro,\\
 17.&&u\to e^{\lambda\omega t^2}\lo u\cos(2\omega t)+v\sin(2\omega t)\ro,\\
&&v\to e^{\lambda\omega t^2}\lo v\cos(2\omega t)-u\sin(2\omega t)\ro,\\
18.&&u\to e^{\nu\omega t}u,\ v\to e^{\nu\omega t}(v-\s\omega tu),\\
19.&&u\to e^{\lambda\omega t^2}u,\ v\to e^{\lambda\omega t^2}(v+2\omega tu),\\
20.&&u\to e^{2\omega t}u,\ v\to e^{\va\omega t^2}v,\\
21.&&u\to u+3\om t,\ v\to v+3\om t^2u+3\om^2t^3+\rho t u+3\omega\rho t^2,\\
22.&&u\to u+\rho x_m,\ v\to v . \ea\ee

Here the Greek letters denote parameters whose values are either
arbitrary or specified in the tables presented below. Equivalence
transformations (\ref{eqv}) are valid only for particular
non-linearities which will be specified in the following.

\section{Symmetries and classifying equations}

We search for symmetries of equations (\ref{1.2}) and (\ref{1.3})
with respect to continuous groups of transformations using the
infinitesimal approach.
Applying the Lie algorithm or its specific formulation proposed in \cite{nikwil1} one can find
the determining
equations for coordinates $\eta ,\ \xi^{a},\ \pi^1,\ \pi^2$ of generator $X$ of the symmetry group:
\begin{equation}
X=\eta {\partial_ t}+\xi^{a}\partial_{x_a}
-\pi ^{1}\partial_ u -\pi ^{2}\partial_{ v}\label{3.105}
\end{equation}
 and classifying equations for non-linearities $f^1$ and $f^2$.
  We will not reproduce the related routine
 calculations but present the general form of symmetry $X$ for equation (\ref{1.2}) with
$a\neq 0, 1$ \cite{nik1} :
 \be \label{2.4} \ba{l}
X=\lbd K+\sigma^\mu G^\mu+\om^\mu \widehat G^\mu+\mu
D-C^{1}u\p_u-C^2v\p_v-B^1\p_{ u}-B^2\p_v\\
+\Psi^{\mu \nu} x_\mu \p_{x_\nu}+\nu \p_t+\rho^\mu\p_{x_\mu} \ea \ee
where the Greek letters denote arbitrary constants, $B^1,\ B^2$ and
$C^1, C^2$ are functions of $t,x$  and $t$ respectively, and
\be\label{2.6} \ba{l} K=2t(t\p_t+x_\mu
\p_{x_\mu})-\frac{x^2}{2}(u\p_u+\frac{1}{a}v\p_v) -tm(u\p_u+v\p_v)
,\\D=t\p_t+\frac12 x_\mu \p_{x_\mu},\\
G_\mu=t\p_{x_\mu}-\frac{1}{2}x_\mu(u\p_u+\frac{1}{a}v\p_v),
\\
\widehat G_\mu=e^{\gm t}\left(\p_{x_\mu}-\frac{1}{2}\gm
x_\mu(u\p_u+\frac{1}{a}v\p_v)\ro . \ea \ee

    If $a=0$ then the related generator $X$ again has the form (\ref{2.4}) where however
    $\lambda=\s^\mu=\om^\mu=C^2=0$ and $B^2$ is a function of $t,x$ and $u$.

    For $a=1$ the symmetry group generator (which we denote by $\tilde X$) is more extended and has
    the following form:
    \be\label{2.41}
    \tilde X=X+C^3u\p_v+C^4v\p_u\ee
    where $X$ is given in (\ref{2.4}) and $C^3,\ C^4$ are functions of $t$.

Equation (\ref{1.2}) admits symmetry (\ref{2.41}) iff the following classifying equations
for $f^1$ and
$f^2$ are satisfied \cite{nik1}:
\be \label{2.7} \ba{l} \left(\lbd(m+4
)t+\mu+\frac{1}{2}\lbd x^2+\sigma^\mu x_\mu+\gm e^{\gm t}
\om^\mu x_\mu)+C^{1}\right) f^1+C^4f^2
+C^{1}_t u\\+C^4_tv +B_t^1-\Delta_m B^1
=\lo B^1\p_u+B^2\p_v+C^{1}u\p_ u+C^2v\p_v +C^3u\p_v+C^4v\p_u\right.\\\left.+\lbd mt(u\p_u+v\p_v)
+\left(\frac12\lbd x^2+\sigma^\mu x_\mu
+\gm e^{\gm t}\om^\mu
x_\mu\right) (u\p_u+\frac{1}{a}v\p_v)\ro f^1,\\\\
\left(\lbd(m+4
)t+\mu+\frac{1}{a}\lo\frac{1}{2}\lbd x^2+\sigma^\mu x_\mu+\gm e^{\gm t}
\om^\mu x_\mu\ro+C^{2}\right) f^2+C^3f^1+C^{2}_t v\\
 +C^3_tu+B_t^2-\Delta_m B^2
=\lo B^1\p_u+B^2\p_v+C^{1}u\p_ u+C^2v\p_v +C^3u\p_v+C^4v\p_u\right.\\\left.+{\lbd mt}(u\p_u+v\p_v)+
\left(\frac12\lbd x^2+\sigma^\mu x_\mu
+\gm e^{\gm t}\om^\mu
x_\mu\right) (u\p_u+\frac{1}{a}v\p_v)\ro f^2.
 \ea \ee
In other words, to make group classification of systems (\ref{1.2}) means to
find all non-equivalent solutions of equations (\ref{2.7}) and to specify the related
symmetries (\ref{2.4}) \cite{nikwil1}. We note that equations (\ref{2.7}) can be decoupled
equating terms multiplied by the same variables $x_\mu$ or their powers.

Consider now equation (\ref{1.3}) and the related symmetry
operator (\ref{3.105}). The determining equations for $\eta,\
\xi^\mu$ and $\pi^a$ are easily obtained using
the standard Lie algorithm:
\be \label{2.8}\ba{l} \eta_{tt}=\eta_{x_\mu}=\eta_u=\eta_v=0,
\\
\xi^\mu_t=\xi^\mu_u=\xi^\mu_v=0, \\\pi^1_v
=\pi^2_u=0,\ \pi^1_{uu}=p\pi^2_{vv}=0;\\
 \pi^a_{x_\mu u}+\pi^a_{x_\mu v}=0,  \\
p(\pi^1_u-\pi^2_v-\frac{1}{2} \eta_t)=0,\\ \xi^\mu_{x_\nu}+\xi^\nu_{x_\mu}=-\delta^{\mu \nu}
\eta_t,\
\mu\neq m \ea \ee
where subscripts denote derivatives w.r.t. the corresponding independent variable, i.e.,
 $\eta_t=\frac{\p\eta}{\p t},\ \xi^\mu_{x_\nu}=\frac{\p\xi^\mu}{\p x_\nu}$, etc.

Integrating system (\ref{2.8}) we obtain the general form of
operator $X$: \begin{gather} \label{2.9} X=\nu \p_t+\rho_\nu
\p_{x_\nu}+\Psi^{\s  \nu}\p_{\nu} x_\s+\mu D-B^1\p_ u-B^2\p_v-Fu\p_u
-G v\p_v;\\ \label{2.90} \mu=2(F-G)\ \  {\rm if}\
p\not=0\end{gather} where $B^1, \ B^2$ are functions of $(t,x)$, $F$
and $G$ are functions of $t$ and summation over the indices $\s,
\nu$ is assumed with $\s, \nu=1,2, \cdots, n-1$.

The classifying equations for $f^1$ and $f^2$ reduce to the
following system \begin{gather} \label{2.10}
(\mu+F)f^1+F_tu+(\p_t-\Delta_m)B^1=\left(B^1\p_u+B^2\p_v+
Fu\p_u+Gv\p_v\right)f^1,\\\label{2.101}(\mu+G)f^2+G_tv+B^2_t-pB^1_{x_m}=\left(B^1\p_u
+B^2 \p_v+F u\p_u+G v\p_v\right)f^2 .\end{gather}

Solving (\ref{2.10}), (\ref{2.101}) we  shall specify both the
coefficients of infinitesimal operator (\ref{2.9}) and the related
non-linearities $f^1$ and $f^2$.

It is obvious that the widest spectrum of symmetries corresponds to the case when the parameter $a$
is equal to 1 since  the corresponding generator $\tilde X$ (\ref{2.41})
includes two additional terms $C^3u\p_v$ and $C^4v\p_u$. Quite the contrary, equations
(\ref{1.2}) with $a\neq 1$ and especially
(\ref{1.3}) admit relatively small variety of symmetries.

\section{Classification of symmetries}

Following \cite{nik1} we specify basic, main and extended symmetries for the analyzed systems of
reaction-diffusion equations.

Basic symmetries are nothing but generators of transformations (\ref{2.2}) forming the kernel of a symmetry
group, i.e.,
\be \label{4.1} P_0=\p_t, \quad
P_\mu=\p_{x_\mu}, \quad J^{\mu \nu}=x_\mu\p_{x_\nu}-x_\nu \p_{x_\mu}. \ee

Main symmetries form an important subclass of general symmetries (\ref{2.4}) and have the following form
\be
\label{4.2} \tilde X=-\mu D+C^{1}u\p_{
u}+C^2v\p_v+C^3u\p_v+C^4v\p_u+B^1\p_{ u}+B^2\p_v \ee
(if $a\neq1$  then $C^3=C^4=0$).

In accordance with the analysis present in \cite{nik1} the complete description of general symmetries
(\ref{2.4}) can be obtained using the following steps:
\begin{itemize}
\item  Find all main symmetries (\ref{4.2}), i.e., solve equations (\ref{2.7}) for
$\Psi^{\mu\nu}=\nu=\rho^\nu=\sigma^\nu=\om^\nu=0$:
 \be \label{4.3}\ba{l} (\mu
+C^{1})f^1+C^4f^2+C^{1}_t u+C^{4}_t v+B^1_t-\Delta_m B^1\\=
(C^{1}u\p_u+C^2v\p_v+C^3u\p_v+C^4v\p_u+B^1\p_u+B^2\p_v) f^1,\\\\
(\mu
+C^{2})f^2+C^3f^1+C^{2}_t v+C^3_tu+B^2_t-a\Delta_m B^2\\=
(C^{1}u\p_u+C^2v\p_v+C^3u\p_v+C^4v\p_u+B^1\p_u+B^2\p_v) f^2.\ea \ee
\item  Specify all cases when the main symmetries can be extended, i.e., at least one
of the following systems be satisfied:
\be
\label{4.8}\ba{l} af^1=(au\p_u+v\p_v)f^1,\\ f^2=(au\p_u+v\p_v)f^2,
\ea\ee \be\label{4.10}\ba{l} a(f^1+\gm u)=(au\p_u+v\p_v)f^1,\\ f^2+\gm
v=(au\p_u+v\p_v)f^2\ea \ee or if equation (\ref{4.8}) is
satisfied together with the following conditions:
 \be\label{4.9}\ba{l}(m+4)f^c+\mu^{c1}f^1+\mu^{c2}f^2\\=
 \lo(\mu^{11}u+\mu^{12}v+mu)\p_{u} +(\mu^{21}u+\mu^{22}v+mv)\p_{v}\ro f^c,
\\\\\nu^{c1}f^1+\nu^{c2}f^2+\mu^{c1}u+\mu^{c2}v=\\\lo(\nu^{11}u+\nu^{12}v)
\p_{u}+ (\nu^{21}u+\nu^{22}v)\p_v\ro f^c\ea\ee
were $c=1,2$,  $\mu^{cb}$ and $\nu^{cb} $
are constants satisfying $(a-1)\mu^{cb}=(a-1)\nu^{cb}=0$.

If relations (\ref{4.8}), (\ref{4.10}) or (\ref{4.9}) are valid then the system (\ref{1.2})
admits symmetry $G^\al, \widehat G^\al$ or the conformal symmetry
$K-(t\mu^{11}+\nu^{11})u\p_{u}-(t\mu^{12}+\nu^{12})v\p_{u}-
(t\mu^{21}+\nu^{21})u\p_{v}-(t\mu^{22}+\nu^{22})v\p_{v}$
 correspondingly.

\item When classifying equations (\ref{1.3}) for $p\neq0$ the second step in not needed
 since in accordance with
(\ref{2.9}) these equations admit only basic and main symmetries.
\end{itemize}

In the following sections we find main and extended symmetries for the classified equations.
For clarity we start with group classification of systems (\ref{1.3}) with $p\neq 0$
 which is more simple
technically and present rather detailed calculations. Then we consider equations
(\ref{1.2}) and present classification results without technical details.

\section{Algebras of main symmetries for equation (\ref{1.3})}

To describe main symmetries we use the trick discussed in \cite{nik1}, i.e., make {\it a priori}
classification of low dimension algebras of these symmetries.
In accordance with (\ref{2.9}) any symmetry generator extending
algebra (\ref{4.1}) has the following form \be \label{5.1} X=\mu
D-B^1\p_u-B^2\p_v-F u\p_u+\lo\frac{\mu}2-F\ro v
\p_v. \ee

Let $X^1$ and $X^2$ be operators of the form (\ref{5.1}) then the
commutator $[X^1, X^2]$ is also a symmetry whose general form is
given by (\ref{5.1}). Thus operators (\ref{5.1}) form a Lie
algebra which we denote as $\cal A$.

Let us specify algebras $\cal A$ which can appear in our
classification procedure. First consider one-dimensional $\cal A$
 , i.e., suppose that equation (\ref{1.3}) admits the only symmetry
of the
form (\ref{5.1}). Then any commutator of operator (\ref{4.1}) with
(\ref{5.1}) should be equal to a linear combination of
operators (\ref{4.1}) and (\ref{5.1}). Using this condition we come to
the following possibilities only: \be \label{5.2} \ba{l} X=X^1=\mu
D- \al^1 \p_{ u}-\al^2 \p_{ v}-\bt u \p_u-(\bt-\frac{\mu}2)
v
\p_v,\\
X=X^2=e^{\nu t}(\al^1 \p_u+\al^2 \p_v+\bt u \p_u+\bt v\p_v),\\
X=X^3=e^{\nu t+\rho \cdot x} (\al^1\p_ u+\al^2\p_ v) \ea \ee where
the Greek letters again denote arbitrary parameters and $\rho\cdot
x=\rho^\mu x_\mu$.

The next step is to specify all non-equivalent sets of arbitrary
constants in (\ref{5.2}) using the equivalence transformations
(\ref{x.2}).

If the coefficient for $u\p_u$ (or $v\p_v$) is
non-zero then translating $u$ (or $v$) we reduce to zero the related
coefficient $\al^1$ ($\al^2$) in $X^1$ and $X^2$; then scaling $u$ ($v$) we can
reduce to $\pm 1$ all non-zero $\al^a$ in (\ref{5.2}). In
addition, all operators (\ref{5.2}) are defined up to constant
multipliers. Using these simple arguments we come to the following
non-equivalent versions of operators (\ref{5.2}):
 \be \label{5.3} \ba{l}
X_1^{(1)}=2\mu D-u \p_u+(\mu-1)v \p_v
,\\
X_1^{(2)}=2D+v\p_v+\nu\p_
u,\
X_1^{(3)}=2D-u \p_u-\p_v,\\
X_2^{(\nu)}=e^{\nu t+\rho\cdot x}(u\p_u+v
\p_v),\\
X_3^{(1)}=e^{\sigma^1t+\rho^1\cdot
x}({\p_{u}}+{\p_{v}}),\
X_3^{(2)}=e^{\sigma^2t+\rho^2\cdot x}{\p_{u}},\
X_3^{(3)}=
e^{\sigma^3t+\rho^3\cdot x}{\p_{v}}. \ea \ee

To describe {\it two-dimensional} algebras $\cal A$ we represent
one of the related basis element $X$ in the general form
(\ref{5.1}) and calculate the commutators
\[
Y=[P^0, X]-2\mu P^0,\ \  Z=[P^0, Y],\ \  W=[X,Y]
\]
where $P^0$ is operator given in (\ref{4.1}).  After simple
calculations we obtain \be \label{5.4} \ba{l}
Y=F_t(u\p_u+v \p_v)+B^1_t \p_u+B^2_t \p_v
,\ \ Z= F_{tt} (u\p_u+v
\p_v)+B^1_{tt}{\p_ u}+B^2_{tt}
{\p_ v}
,\\
W=2\mu t Z+\mu x_b (B^1_{tx_b}{\p_ u}+B^2_{tx_b}{\p_v}). \ea \ee

By definition, $Y$, $Z$ and $W$ belong to $\cal A$. Let $F_t\ne 0$
then we obtain from (\ref{5.4}): \begin{gather} \label{5.5}\mu\neq
0:\ B^a_{tt}=F_{tt}=B^a_{tb}=0, \\ \label{5.6} \mu=0: \ F_{tt}=\al
F_t+\gm^a B^a_t, \quad B^a_{tt}=\gm^aF_t+\bt^{ab} B^b_t.
\end{gather}

Starting with (\ref{5.5}) we conclude that up to translations of
$t$ the coefficients $F$ and $B^a$  have the following form
\[
F=\sigma t \ \mbox{or} \ F=\bt;\  B^a=\nu^a t+\al^a \ \mbox{if} \
\mu \ne 0.
\]

If $F=\sigma t$ then the change \be\label{5.7} u_a \to u_a
e^{-\sigma t}-\frac{\nu^a}{\mu} t\ee reduces the related operator
(\ref{2.9}) to $X^1$ of (\ref{5.2}) for $\bt=0$.

The choice $F=\bt$ corresponds to the following operator
(\ref{5.1}) \be \label{5.9} X=X^4=X^1-2t( \al^1 {\p_ u}+\al^2 {\p_ v})
\ee where $X^1$ is given in (\ref{5.2}).

Thus if one of basis elements of two dimension algebra $\cal A$ is
of general form (\ref{5.1}) with $\mu \ne 0$ then it can be reduced
to $X^1$ with $\bt=0$ or to generator (\ref{5.9}). We denote such
basis element as $e^1$. Without loss of generality the second basis
element $e^2$ of $\cal A$ is a linear combination of operators
$X_2^{(\nu)}$ and $X_3^{(a)}$ (\ref{5.3}). Going over possible pairs
$(e^1, e^2)$ and requiring $[e^1, e^2]=\al^1 e^1+\al^2 e^2 $ we come
to the following two dimensional algebras:
 \be \label{5.10} \ba{l}
A_1=<2D+v\p_v, X_2^{(0)}>, \quad A_2=<X_1^{(2)},
X_3^{(3)}>,\\
A_3=< X_1^{(3)}, X_{(3)}^3>, \quad A_4=<X_1^{(1)},\quad
X_3^{(3)}>,\\
A_5=<X_{(1)}^1, X_3^{(3)}>, \quad A_6=<2D+2 v {\p_
v}+u
\p_u+\nu t \p_v, X_3^{(2)}>,\\
A_7=<2D+2 u \p_u+3 v \p_v+3\nu
t\p_u, \quad X_{(1)}^3>. \ea \ee The form of basis
elements in (\ref{5.10}) is defined up to transformations (\ref{x.2}),
(\ref{5.7}).

If $\cal A$ does not include operators (\ref{5.1}) with
non-trivial parameters $\mu$ then in accordance with (\ref{5.7})
its elements are of the following form \be \label{5.11}
e_a=F^{(a)}\left(u\p_u+v{\p_
v}\right)+B^1_{(a)} {\p_ u}+B^2_{(a)} {\p_ v}, \ a=1,2 \ee where
$F^{(\al)}$
and $B^1_{(a)}$, $B^2_{(a)}$ are solutions of (\ref{5.6}).

Formulae (\ref{5.10}), (\ref{5.11}) define all non-equivalent
two-dimensional algebras $\cal A$ which have to be considered as
possible symmetries of equations (\ref{1.3}). We will see that
asking for invariance of (\ref{1.3}) w.r.t. these algebras the
related arbitrary functions $f^a$ are defined up to arbitrary
constants, and it is impossible to make further specification of
these functions by extending algebra $\cal A$.

\section{ Group classification of equations (\ref{1.3})}

We suppose that parameter $p$ in (\ref{1.3}) be nonzero. Then,
scaling independent variables $t, x$ we can reduce it to $p=1$.

To classify equations (\ref{1.3}) which admit one- and two-
dimension extensions of the basis invariance algebra (\ref{4.1})
it is sufficient to solve determining equations (\ref{2.10}) for
$f^a$ with {\it known} coefficient functions $B^a$ and $F$ of
symmetries (\ref{5.1}). These functions are easily found comparing
(\ref{2.9}) with (\ref{5.3}), (\ref{5.10}) and (\ref{5.11}).

Let us present an example of such calculation which corresponds to
algebra $A_1$ whose basis elements are $X^1=2t\p_t+x_a \p_{x_a}+
v\p_v$ and $X_2^{(0)}=u \p_u+
v\p_v$, refer to (\ref{5.10}). Operators $X^1$ and $X_2^{(0)}$
generates the following equation (\ref{2.10}), (\ref{2.101}):
\be\label{6.2}f^1=-u
f^1_u; \quad f^2=-\frac{1}{2} u f^2_u\ee
and
\be \label{6.21}f^a=\left(u\p_u+v\p_v\right)f^a,
\quad a=1,2.
\ee

General solution of (\ref{6.21}) is: $
f^1=uF^1\left(\frac{v}{u}\right), \quad
f^2=uF^2\left(\frac{v}{u}\right)$ were $F^1$ and
$F^2$ are arbitrary functions of $\frac{v}{u}$. Solving (\ref{6.2})
for such functions
$f^1$ and $f^2$ we obtain
\be \label{6.3}
 f^1=\al u^3 v^{-2}, \quad f^2=\lbd u^2  v^{-1}.
\ee

Thus equation (\ref{1.3}) admits symmetries $X^{(2)}_0$ and $X^1$
provided $f^1$ and $f^2$
are functions  given in (\ref{6.3}). These symmetries are defined up
to arbitrary constants $\al$ and $\lbd$. If one of these constants
is nonzero, than
it can be reduced to $+1$ or $-1$ by scaling independent variables.

In analogous way we solve equations (\ref{2.10}) corresponding to
other symmetries presented in (\ref{5.3}) and (\ref{5.10}). At that we do not
consider functions $f^1$ and $f^2$ which are ether linear in $u,\ v$ or correspond
to decoupled systems (\ref{1.3}) (i.e., when $f^1$ and $f^2$ depend only on $u$ and $v$
correspondingly).
The classification results are presented in
Table 1.

In the fourth column of the table symmetries of the related equation
(\ref{1.3}), {\ref{1.31}) are presented together with the additional
equivalence transformations (AET) which are listed in formula
(\ref{eqv}); the numbers of AET from the list (\ref{eqv}) are given
in square brackets.
 Greek letters denote arbitrary real parameters which in particular can
 be equal to zero. Moreover, without loss of
generality we restrict ourselves to $\eta=0,1$, $\delta=0, \pm 1, \varepsilon=\pm1$.

In Table 1 $D$ is the dilatation operator given in (\ref{2.6}),
$\tilde x=(x_1, x_2, \cdots, x_{m-1}),$ $\Psi(x)$ is an arbitrary
function of spatial variables; $ \Psi_\mu(\tilde x,x_m-\eta t)$ and
$ \Phi_\mu(t, \tilde x) $ are solutions of the Laplace and linear
heat equations:
$$\ba{l} \Delta_{m} \Psi_\mu=\mu
\Psi_\mu, \  \ (\frac{\p}{\p t}-\Delta_{m-1} )\Phi_\mu
=\mu\Phi_\mu.\ea$$
\begin{center}{\bf
Table 1. Non-linearities and symmetries for equations (8), (9) with
p=1}
\end{center}
\begin{tabular}{|l|l|l|l|}
\hline No& $ \mbox{Non-linearities} $ & ${\ba{l}\mbox{Arguments}\\
\text{ of}\ F^1,F^2 \ea}$ & $\ba{l}\mbox{Symmetries}
\\\texttt{and AET Eq.(15)}\ea$\\
  \hline 1 & $\ba[t]{l}f^1=u^{2\nu+1} F^1\\f^2=u^{\nu+1} F^2\ea$, &
  $v u^{\nu-1}$ &
$\ba{l}2\nu
D-u\p_u+(\nu-1)v\p_v\\{[}\texttt{AET 1 }\texttt{if }\nu=0{]}\ea$\\
 \hline 2 & $\ba{l}f^1=F^1 v^{-2},\\  f^2= F^2 v^{-1}\ea$ & $u-\eta \ln v$ &
$\ba[t]{l}
2D+v\p_v+\eta \p_u\ea$\\
\hline 3 & $\ba{l} f^1=u(F^1+\varepsilon \ln u),
\\f^2=v(F^2+\varepsilon \ln u)\ea$ & $\frac{v}{u}$ &
$\ba{l}e^{\delta t}\left( u\p_u+ v{\p_v
}\right)\ea$\\

\hline
4 & $\ba{l}f^1=u^3F^1,\\ f^2=u^2F^2\ea$ & $v-\ln u$ & $\ba{l}2D-u{\p_
u}-
\p_v\ea$ \\
  \hline
  5 &  $\ba{l}f^1=F^1+\mu u,\\ f^2=F^2+\eta u\ea$  &  $v$  &
  $\ba{l}e^{-\eta x_m}
  \Phi_\mu(t,\tilde x){\p
_u}\\ {[}\texttt{AET 2, 22 if }\eta=\mu=0{]}\ea$\\
\hline
 6 &  $f^1=F^1+\eta(\mu-\nu)v $  &  $u+\eta v$  &  $e^{\nu t}
\Psi_\mu(\tilde x, x_m-\eta t)(
\p_v-\eta \p_u)$, \\
 & $f^2=F^2+ \nu v$  &   &$\ba{l}
 {[}\texttt{AET 11 if }\nu=\mu=0{]}\ea$
\\
\hline  7 & $\ba{l}f^1=\delta u^3 v^{-2}, \\ f^2=\mu u^2 v^{-1}\ea$ &&
$\ba{l}2D+v \p_v, \quad u{\p_u}+v{\p_
v}\\{[}\texttt{AET 1 }{]}\ea$\\
\hline 8 & $\ba{l}f^1=\nu e^{-2 u}, \\ f^2=\eta e^{-u} \ea$ &&
$\ba{l}2D+v \p_v+\p_u, \
\Psi(x) \p_v\\\texttt{[AET 3]}\ea $\\
\hline 9 & $f^1=\eta e^{3 v},$ && $2D-u {\p_ u}-{\p_ v},\
\Phi_0(t,\tilde x)
\p_u$ \\
  &  $f^2=\delta e^{2 v}$  &&  $\ba{l}{[}\texttt{AET 2, 22}{]}\ea$ \\
\hline 10 & $\ba{l}f^1=\mu u^{2\nu+1}, \\f^2=\eta u^{\nu+1}\ea$ &&
$\ba{l}2\nu D-u\p_u+(\nu-1)v \p_v, \ \Psi(x)\p_v\\{[}\texttt{AET
3}{]}\ea$\\
\hline 11 & $f^1=\eta v^{3\nu-2},$ && $2(\nu-1)D-\nu u \p_u-v \p_v,\
\Phi_0(t,\tilde x)
\p_u$ \\
 & $f^2=\mu v^{2\nu-1}$  &&  $\ba{l}{[}\texttt{AET 2, 22; }{]}
  \ea$\\
\hline 12 & $f^1=\frac{\nu}{u}, \quad f^2=\ln u$ &&
$\ba{l}2D+2v\p_v+u\p_u+t{\p_ v},
\  \Psi(x)\p_v\\{[}\texttt{AET 3; }\&\ 7\ \texttt{ if }\nu=0{]}\ea$\\
\hline 13 & $f^1=\ln v, \quad f^2=\nu v^{\frac{1}{3}} $  &&
$\ba{l}2D+2 u\p_u+3 v \p_v+3 t\p_u, \\  \Phi_0(t,\tilde x)
\p_u\\  {[}\texttt{AET 2, 22; }\&\ 8\ \texttt{ if }\nu=0{]}\ea$\\
\hline
\end{tabular}


\section{Group classification of equations (\ref{1.2})}

In this Section we present the classification results for
coupled systems of equations (\ref{1.2}).
The related classifying equations are
given by relations (\ref{2.7}).

Like in Section 5 we first describe all non-equivalent low dimension
algebras of the main symmetries for equation (\ref{1.2}). Non-equivalent realizations of
these algebras  (together
with detailed calculations) are present in the Appendix.
Using found realizations of algebras $\cal A$ and solving the related classifying
equations (\ref{4.3}) we easily complete the group classification of
equations (\ref{1.2}).

We will not reproduce here the related routine  calculations but present
the results of group classification in Tables 2-10. Besides symmetries and the related non-linearities, the additional
equivalence transformations which are admissible by
particular classes of equations (\ref{1.2}) are indicated there.
The symbols  $D, \ G^\mu, \ \widehat G^\nu$ and $K$ denote
generators listed in (\ref{2.6}),
 $\psi_\mu$, $\tilde \psi_\mu$ and $\Psi_\mu=\Psi_\mu(x)$ are arbitrary solutions of the
linear heat equations and Laplace equation:
$$\ba{l}\p_{ t}\psi_\mu-\Delta_m \psi_\mu=\mu\psi_\mu,\
\p_{ t}\tilde\psi_\mu-a\Delta_m \tilde\psi_\mu=\mu\tilde\psi_\mu,\  \Delta_m
\Psi_\mu=\mu\Psi_\mu,\ea$$
and $\Psi(x)$ is an arbitrary function of $x$.
The Greek letters denote arbitrary parameters.  Moreover, up to equivalence transformations
we restrict ourselves to $\va=\pm1$, $\eta=0,1$ and $\delta=0, \pm1$ .

\begin{center}
{\bf Table 2.
Non-linearities with arbitrary functions and symmetries
for equations (2) with arbitrary
$a\neq0$}
\end{center}
\begin{tabular}{|c|c|c|c|c|}
  \hline
  No&Nonlinear terms&\begin{tabular}{l}
Arguments\\ of \
$F^1$, $F^2$ \\
\end{tabular}&Symmetries&$\ba{l}
\texttt{AET}\\ \texttt{Eq.}(15)\ea$
\\
  \hline

 1& $\begin{array}{l}
f^1=u^{\nu +1}F^{1}, \\
f^2=u^{\nu -\mu }F^2
\end{array}$
& ${{v}{u^\mu }}$ &$\begin{array}{l} \nu D-u{{\partial_u}}+\mu
v{{\partial_v}}\\\texttt{for any}\ \mu,\nu,\\ \& \ G_\al\
\texttt{for}\ \nu=0,\\ a\mu=1 \end{array}$&$\ba{c}14,
\\ \rho=\mu\omega\\\texttt{if}\ \nu=0\ea$\\
\hline
2 &$
\begin{array}{l}
f^1=u(F^1+\va \ln u),
 \\
f^2=v(F^2+\va \ln v)
\end{array}$
&$ {{v}{u^\mu }} $& $\ba{l} e^{\va t}\left(
u{{\partial_u}}-\mu v{{\partial_v}}\right)\\\texttt{for any }\ \mu, \ \&\\
\widehat G_\al\ \texttt{if}\ a\mu=-1\ea$&\\\hline
3 & $\ba{l} f^1=v^\nu F^{1}, \\ f^2=v^{\nu +1}F^2 \ea$&$
u-\ln v $&$\ba{l} \nu D-v{{\partial_v}}-{{\partial_u}}\ea$&$\ba{c}
4, \mu=-1\\ \text{ if}\ \nu=0 \ea $\\
\hline 4 &$ \ba{l}f^1=F^1+\va u,\\ f^2=F^2v+\va uv \ea $&$ u-\ln
v $&$ e^{\va t}\left( v{{\partial_v}}+{
{\partial_u}}%
\right)  $&\\
    \hline
 $5$& $\ba{l}f^1=0, \  \ f^2=F^2,\\a\neq1\ea$&$u$&$\ba{l}D+v\p_v,\
 \tilde\psi_0\p_v\ea$
    &$\ba{c}  3\ea$\\
\hline
6 &$\ba{l} f^1=F^1,\\ f^2=F^2+\delta v \ea$&$ u $&$\ba{l}
\tilde\psi _\delta {\partial_v}\ea
$&$\ba{c}3 \ \texttt{if}\
\delta=0; \\
6\ \texttt{if} \  a=1,\\F^1=\delta u \ea$
\\
\hline 7&$
\begin{array}{l}
f^1=F^1+\delta u, \\
f^2=F^2+\s v,\\  a\neq1
\end{array}
$&$ v- u $&$\ba{l} e^{\kappa t}\Psi _\mu(x) \left({\partial_u}+
{\partial_v}\right),\\ \mu=\frac{\s-\delta}{(1-a)},
\\\kappa=\s+a\mu\ea $&\\
\hline 8 &$
\begin{array}[t]{l}
f^1=e^{ u}F^1, \\  f^2=e^{ u}F^2\\\eta=0\ \texttt{if}\  a=1
\end{array}$
& $\ba{l} v-\eta u,\ea$ & $ D- {\partial_u}-\eta{\partial_v}$ & \\
\hline
\end{tabular}

 \vspace{2mm}

 In Table 3
$\Delta $ denotes the characteristic determinant,
$\Delta=\frac 14(\mu-\nu)^2+ \lambda\sigma.$
Additional equivalence transformations are specified in the third column.
\begin{center}
\textbf{Table 3.  Symmetries of equations (2) with arbitrary $a\neq0$ and
non-linearities $f^1=u\lo \mu\ln u+\lambda \ln v\ro,\ f^2=v\lo
\nu\ln v+\sigma \ln u\ro$}
\end{center}
\begin{tabular}{|l|l|l|l|}
\hline
No&Conditions &Symmetries and AET Eq.(15)&Additional symmetries\\
\hline $1$&$\ba{l}\lambda=0, \s=\va,\\ \mu=\nu\ea$&$\ba{l} e^{\mu
t}v{\partial_v},\ e^{\mu t}\lo u {\partial_u} +\va t
v{\partial_v}\ro\\{[}\texttt{AET 20 if}\ \mu=0 {]}\ea$&none\\
 \hline
$2$&$\ba{l}\lambda=0, \s=\va\\\mu\neq \nu,\
(a-1)^2\\+\nu^2\neq0\ea$&$\ba{l}e^{\mu t}\lo\lo \mu-\nu\ro u
{\partial_u}+ \sigma v {\partial_v}\ro,\\ e^{\nu t}v{\partial_v}\
{[}\texttt{AET 14,}\omega=-\va\nu\rho\\ \texttt{if }\ \mu\nu=0{]}\ea
$ &$\ba{l}
G_\al\ \text{if}\ \nu=-a\sigma,  \mu=0;\\
\widehat
G_\al\ \text{if}\ \mu\neq 0,  \mu-\nu=a\sigma
\ea
$\\
\hline $3$&$\ba{l}\Delta=0,
 \lambda\s\neq 0,
\ea$&$\ba{l}X_2= e^{\Omega t}( 2\lambda u{\partial_u}+( \nu-\mu)
v{\partial_v}),\ea$&$\ba{l}
G_\al\ \text{if} \ \mu=-\nu, \lambda =a\nu\ea$\\
 &$\ba{l}\lambda^2+\s^2=1\\ \mu+\nu=2\Omega
\ea$&$\ba{l}\ 2e^{\Omega t} v{\partial_v}+tX_2\ {[}\texttt{AET
14,}\\
\sigma\omega=-\nu\rho \texttt{ if } \mu+\nu=0{]}\ea$&$\ba{l}\widehat
G_\al\ \text {if} \ \nu\neq -
\mu,\\2\lambda =a(\nu-\mu)\ea$\\
 \hline
 $4$&$\ba{c}\lambda\sigma\neq 0,\ea$&
 $X_\pm=e^{\omega_\pm t}\lo \lambda u{\partial_u}+\lo \omega_\pm-\mu\ro
v{\partial_v}\ro$&$\ba{l}G_\al\
\text{if} \ \nu\mu=\lambda\s, \lbd=-a\mu\ea$\\ &$\ba{l}\Delta=1,\\
\omega_\pm=\Omega\pm 1\ea $ & $\ba{l}{[}\texttt{AET 14, }
\sigma\omega=-\nu\rho\\ \texttt{if} \ \mu\nu=\lambda\sigma{]}\ea$&
$\ba{l}\widehat G_\al\ \text{if}
\ \mu\nu\neq\lbd\s,\\
\lambda=a(\nu-\mu+a\s);\ea$\\
 \hline $5$&$\ba{l}\Delta=-1\ea $&$  \ba{l}e^{\Omega t}\lo
2\lambda\cos tu{\p_u}+\right.\\\left.\lo(\nu-\mu)\cos
t-2\sin
t\ro v{\partial_v}\ro, \\
e^{\Omega t}\lo 2\lambda\sin
tu{\p_u}+\right.\\\left.\lo(\nu-\mu)\sin
t+2\cos t\ro v{\partial_v}\ro
   \ea   $& none\\
\hline
\end{tabular}
\begin{center}{\bf
Table 4. Non-linearities with arbitrary parameters and extendible
symmetries for equation (2) with any $a$}
\end{center}
\begin{tabular}{|l|l|l|l|l|}
\hline No & \text{Nonlinear terms} &$\ba{l}\text{Main}\\
\text{symmetries}\ea $&$
\begin{array}{l}
\text{Additional}\\\text{symmetries}\ea$&$\ba{l}\text
{AET} \\\texttt{Eq.(15)}\ea$\\
\hline
 1 &$
\begin{array}{l}
f^1=\varepsilon u^{\nu +1}v^\mu , \\
f^2=\sigma u^\nu v^{\mu +1},\\a\neq0
\end{array}
$&$
\begin{array}{l}
\mu D-v{{\partial_v},} \\
\nu D-u{{\partial_u}}
\end{array}
$&$
\begin{array}{l}
G_\al\text{ if }\ a\nu =-\mu,\\ \text{ \&
 }\
K\ \text{if }\\ \nu{m(1-a)} =4 ;
\end{array}$&$\ba{l}14,\\\nu\om+\mu\rho\\=0\ea$\\
\cline{4-5} &&&$\ba{l} \psi _0{\partial_u}\ \text{ if }\ \sigma
=0,\\ \nu=-1,\ \text{\&}\  G_\alpha\\ \text{if}\  \mu =a, \  \text{\&}\  K\\ \text{if}\ a=1+\frac {m}4;
\ea$&$\ba{c}2;\ 14,\\\om=\mu\rho\ea$\\
\hline $2$ &$
\begin{array}{l}
f^1=\va u^{\nu +1}, \\
f^2= u^{\nu +\mu },\\\nu^2+(a-1)^2\neq0
\end{array}
$&$
\begin{array}{l}
\nu D-u{{\partial_u}}-\mu v{{\partial_v},}\\ \tilde\psi _0{\partial_v}
\end{array}
$&$\ba{l} G_\alpha\ \text{ if }\ \nu =0,\\ a\mu =1\ea$&$\ba{l} 3;\
\&\ 14,\\\rho=\mu\omega
\\\texttt{if}\ \nu=0\ea$ \\
\hline
    $3$&$\ba{l} f^1=\delta,\\ f^2=\ln u,\ a\neq1
\ea$&$
\begin{array}{l}
 D+u{{\partial_u}}+v{{\partial_v}}{+
t{\partial_v}},\\ \tilde\psi _0 {\partial_v}
\end{array}
$&$
\begin{array}{l}
 u{{\partial_u}+ t
{\partial_v}} \\   \text{ if }\delta =0\ea$ &$\ \ 3, 7,9$\\
\hline
$4$&$
\begin{array}{l}
f^1=\delta u\ln u,\\f^2=\nu v+\ln u,
\end{array}
$&$ \ba{l}\\ \ \tilde\psi_\nu {\partial_v} \ea$&$
\begin{array}{l}
e^{\nu t}\left( u{\partial_u}+t{\partial_v}\right) \\ \text{ if }\nu =
\delta\ea$&$\ba{c}
\ea$ \\
\cline{4-5} &$\ba{l}a\neq1,\\ \nu^2+\delta^2\neq0\\ \ea$&&$\ba{l}
e^{\delta t}\left( (\delta -\nu
)u{\partial_u}\right.\\\left.+ {\partial_v}\right)  \text{ if }\
\nu \neq
\delta
\end{array}
$&$\ba{c} 5, \kappa=\frac1\nu\\ \texttt{if}\ \delta=0\ea$\\\hline
$5$ &$
\begin{array}{l}
f^1=\delta e^{\nu u}, \\
f^2=e^{(\nu +1)u}
\end{array}
$&$\ba{l} \nu D-v{\partial_v}-{{\partial_u}},\\
\tilde\psi _0{\partial_v} \ea$&$\ba{l}(u-\delta t)\p_v\
\texttt{if}\\\nu=0,a=1\ea$& $\ba{l}3,\ \& \   9, \ 4,\\\mu=-1\\
\texttt{if}\ \nu=0\ea$\\
\hline
\end{tabular}

\vspace{3mm}

Classification results present in
 Tables 4 and 5 are related to systems (\ref{1.2}) with arbitrary
 values of $a$ presented by equation (\ref{x.0}) (if not specified in the
 second
 columns of the tables).

 The non-linearities given in Table 4 are defined
up to arbitrary parameters. For some values of these parameters the related
equation (\ref{1.2}) admits extended symmetries indicated in
Column 4 of the table.

In Table 5 non-linearities for equation (\ref{1.2}) are classified whose
symmetries are fixed for all admitted values of parameters.

\begin{center}{\bf
Table 5. Non-linearities with arbitrary parameters and
non-extendible symmetries for equations (2) with arbitrary $a$}
\end{center}
\begin{tabular}{|l|l|l|l|}
\hline No &$ \text{Nonlinear terms} $ &$ \text{Symmetries} $&
$\ba{l}\text{AET Eq.}\ (15) \ea$\\
\hline  $1$ &$\ba{l} f^1=\delta \left( u+v\right) ^{\nu +1},\\
f^1=\mu \left( u+v\right) ^{\nu +1},\ \ a\neq 1 \ea$&$ \ba{l}\nu
D-u{{\partial_u}}-v{\partial_v},\\
\Psi_0 (x)\left( {{\partial_u}-{\partial_v}}\right) \ea
$&$11,\eta=1$\\\hline
$2$ &$
\begin{array}{l}
f^1= e^{v}, \\
f^2=\va e^{v},\ a\neq0
\end{array}
$&$\ba{l} D-{{{\partial_v},}}\  \psi _0{\partial_u}
\ea$&$\ba{c}2
\ea$\\
\hline
$3$ &$\ba{l} f^1=\delta e^{u+v},\\ f^2=\sigma e^{u+v},\ a\neq 1
\ea$&$\ba{l} D-{\partial_v},\\
\Psi_0 (x)\left( {\partial_u}-{{\partial_v}}\right) \ea$&$11,\
\eta=1$ \\
\hline
 $4$ &$\ba{l} f^1=\va v^{\mu }e^{u},\ f^2=\sigma
v^{\mu+1} e^{u},\\ a\neq0,\ \s^2+\mu^2\neq0\ea$&$\ba{l}
D-{\partial_u},\ v{\partial_v}-\mu {{\partial_u}}\ea $&$\ba{c}4,\ \
\text{if}\\
 \s=0
\ea$\\
\hline $5$ &$\ba{l} f^1=\varepsilon e^{ u}, \
f^2= u \ea$&$\ba{l} D+{v}{\partial_v} - {\partial_u}- t
 {\partial_v},\\ \tilde\psi _0{\partial_v} \ea$&$\ \ \ \ 3$\\
\hline
$6$&$\ba{l} f^1=\va\ln \left( u+v\right) ,\\ f^2=\nu \ln \left(
u+v\right),\\a\neq 1\ea$&$
\begin{array}{l}
\Psi_0(x)\lo{\partial_u}- {\partial_v}\ro,\\
\ \va(a-1)\lo D+u{{\partial_u}}+v{{\partial_v}}\ro\\
{+}\left( (a+\va\nu )t\right.\\\left.+\frac{1+\va
\nu}{2m}x^2\right)\lo{\partial_u}- {\partial_v}\ro
\end{array}
$&$11,\ \eta=1$\\
    \hline
$7$&$\ba{l} f^1=\va u^{\nu +1},\  f^2=\ln u,\\
\nu \neq -1\ea $&$\ba{l} \nu \left(
D+{v}{{{\partial_
v}}}\right)-u
{\partial_ u}-t{\partial_v},\\
\tilde\psi
_0 {\partial_v}\ea $&$\ba{c} 3,\ \&\ 7\\  \texttt{if}\ \nu=0\ea$\\
\hline

     $8$&$\begin{array}{l}
f^1=(\mu-\nu) u\ln u+uv, \\
f^2=-\nu^2 \ln u+(\mu+\nu) v
\end{array}
$&$
\begin{array}{l}
X_3=e^{\mu t}\left( u{\partial_u}
+\nu{\partial_v}\right) , \\
tX_3+e^{\mu t}{{\partial_v}}
\end{array}
$&$\ba{c}5, \kappa=-\nu\\\text{if}\ \mu=0 \ea$ \\
    \hline
        $9$&$\begin{array}{l}
f^1=(\mu-\nu) u\ln u+uv, \\
f^2=\lo 1-\nu^2\ro \ln u\\+(\mu+\nu) v
\end{array}
$&$
\begin{array}{l}X_4^\pm=e^{(\mu\pm1)t}(u\p_u\\+(\nu\pm1)\p_v)
\end{array}
$&$\ba{c}5, \kappa=\mu-\nu\\\text{if}\ \mu=\pm 1 \ea$\\
\hline
$10$&$\begin{array}{l}
f^1=(\mu-\nu) u\ln u+uv, \\
f^2=(\mu+\nu) v\\-\lo 1+\nu^2\ro \ln u
\end{array}
$&$
\begin{array}{l}
e^{\mu t}\left( \cos t(u{{\partial_u}}+\nu\p_v)\right.\\\left.-\sin
 t {{\partial_v}}\right), \\
e^{\mu t}\left( \sin  t(u{{\partial_u}}+\nu\p_v)\right.\\
\left. +\cos t {{\partial_v}}\right)
\end{array}$&  \\
\hline
\end{tabular}

\vspace{3mm}

In Tables 6-9 and 10 the additional symmetries are presented which
 correspond
 to the specific values $a=1$ and $a=0$ of the diffusion
 coefficient. We use the following notations here:  $R=\sqrt{u^2+v^2},\
z=\tan^{-1}\frac vu$.

 In Table 6 the AET are given in square brackets and placed
  in the last column.

  \newpage

\begin{center}
{\bf Table 6.
Additional non-linearities with arbitrary functions and symmetries for equations (2) with
$a=1$}
\end{center}
\begin{tabular}{|c|c|c|c|}
  \hline
  No&Nonlinear terms&\begin{tabular}{l}
Arguments\\ of \
$F^1$, $F^2$, $F$ \\
\end{tabular}&$\ba{l}\text{Symmetries}
\\\text{and AET Eq.}\ (15)\ea$
\\\hline
 $1$&$
\begin{array}{l}
f^1=uF^1+\delta \eta v, \\
f^2=\delta\frac{v }{u}(u+\eta v) \\
+uF^2+vF^1,\\\delta^2+\eta\neq0
\end{array}$
& $ue^{-\eta\frac{v}{u}}$ &$
\begin{array}{l}
e^{\delta t}\left(u{{\partial_v}} +\eta (
u{{\partial_u}}+v{{\partial_v}})\right)\\\texttt{ for any }\eta,F_1\\{\&}\
{\tilde\psi}_\al\p_v\ \texttt{for}\  \eta=0,\\F^1=\al-\delta\neq0,
\text{[AET 3} \\ \text{if} \
F^1=-\delta, \ \eta=0{]}
\end{array}$
\\
\hline
    $2$ &$
\begin{array}{l}
f^1=u^{\nu +1}F^1, \\
f^2=u^\nu \left( F^1v+F^2u\right)
\end{array}$
& $ue^{-\eta\frac{v}{u}}$&  $\begin{array}{l}\eta
(\nu D-u{{\partial_u}}-v{
{\partial_v})-u\p_v%
}\\{[}\texttt{AET 6 if}\ \eta=0{]}
\end{array}$
\\
\hline
$3$ &$
\begin{array}{l}
f^1=uF^1+vF^2 \\
+\va z\left( \mu u-v\right),\\f^2=vF^1-uF^2\\+\va z\left( \mu v
+u\right) \ea$ &$\ba{c}R{}e^{-\mu z}
\end{array}$&$
\begin{array}{l}
e^{\va t}\left( \mu R{{\partial_ R}}+{\p_ z}\ro\ea$\\
\hline $4$ &$\ba{l} f^1=e^{\eta u}F^1,\\f^2=e^{\eta
u}(F^2+F^1u)\ea$&$ 2v-u^2 $&$\ba{l} \eta D-u{{\partial_v}}-{{\partial_u}}\\
{[}\texttt{AET 13 if} \ \eta=0{]}\ea$\\
\hline $5$ &$\ba{l} f^1=\va u+F^1,\\ f^2=\va u^2+F^1u+F^2
\ea$&$ 2v-u^2 $&$ e^{\va t}\left( u{{\partial_v}}+{
{\partial_u}}\right)  $\\
\hline
6&$\begin{array}{l}f^1=Fu,\\ f^2=Fv\ea$&$u$&
$\ba{l} v\p_v,\ u\p_v\
\\ \text{[AET }\  6;\ \&\ 14,\omega=0 ]\ea$\\
\hline $7$&$\ba{l} f^1=\eta,\ f^2=\delta v+F,\quad
 \ea$&$ u $&$ \ba{l}{\tilde\psi} _\delta {\partial_v},
\ e^{\delta t}\left(u-\eta t\ro{\partial_v}\\ \texttt{ for any }
\eta,\delta,\
 \&\\ D+v\p_v\
\texttt{for }
\eta=\delta=0\\{[}\texttt{AET 3 if}\ \delta=0\\ \&\ 6 \texttt{ if }\eta=0{]} \ea $\\
    \hline
$8$ &$
\begin{array}{l}
f^1=e^{\lambda z }\left( F^1v+F^2u\right) ,  \\
f^2=e^{\lambda z }\left( F^2v-F^1u\right)
\end{array}
$&$ R\,e^{\nu z } $&$
\begin{array}{l}
\lambda D+\nu \left( u{{\partial_u}}+v{
{\partial_v}}\right)  \\
- u{{\partial_v}}+v{{\partial_u}%
}\\ {[}\texttt{AET 15,} \ \s=1 \ \texttt{if}\
\lambda=0{]}\end{array}$
\\\hline
9. &$
\begin{array}{l}
f^1=e^{{\frac{v}{u}}}F^1u, \\
f^2=e^{ {\frac{v}{u}}}\left( F^1v+F^2\right)
\end{array}$
&$u $& $\begin{array}{l}  D-u{ {\p_v}}\ea$
\\
 \hline $10$ &$\ba{l} f^1= u^2, f^2=(u+\delta) v+F \ea$&$ u
$&$\ba{l} e^{\delta t} u {\partial_v},\  e^{\delta t}\left(
{\partial_v}+tu
{\partial_v}\right)  \ea$\\
\hline
$11$ &$\ba{l} f^1= \left( u^2-1\right),\\
f^2=\left(  u+\nu \right) v+F \ea$&$ u $&$
\begin{array}{l}
e^{(\nu +1 )t}\left( u{{\partial_v}+ {\partial_v}}\right) , \\
e^{(\nu -1 )t}\left( u{{\partial_v}- {\partial_v}}\right)
\end{array}$
\\
\hline
$12$ &$\ba{l} f^1= \left( u^2+1\right) ,\\
f^2=\left(  u+\nu \right) v+F \ea$&$ u $&$
\begin{array}{l}
e^{\nu t}\left( \cos t u{\partial_v}- \sin
t{\partial_v}\right),
 \\
e^{\nu t}\left( \sin tu{\partial_v}+\cos
t{\partial_v}\right)
\end{array}$ \\
 \hline
\end{tabular}

\vspace{5mm}

In the following Tables 7 and 8 $\Delta=\frac 14(\mu-\nu)^2+
\lambda\sigma$.
 Symmetries and additional equivalence transformations are specified in the third
 column; AET are given in square brackets. In the last columns
 additional symmetries are specified which are valid for some particular values
 of parameters defining non-linearities.

\newpage
\begin{center}
\textbf{Table 7.  Symmetries of equations (2) with $a=1$ and
non-linearities $f^1=(\mu u-\sigma v)\ln R+z(\lambda u- \nu v)$,
$f^2=(\mu v+\sigma u)\ln R+z(\lambda v+\nu u)$}
\end{center}
\begin{tabular}{|l|l|l|l|}
\hline
No & Conditions & Symmetries and AET Eq. (15)& Additional\\
& for coefficients &  & symmetries  \\
\hline $1$ & $\ba{l}\lambda =0,\\ \mu =\nu=\delta\ea$ &
$\ba{l}e^{\delta t}{\partial_z },\ e^{\delta t}\lo R{\partial_ R}
+\sigma t{\partial_z}\ro
\\{[}\texttt{AET 16 if}\ \mu=0{]}
\ea$ & $\ba{l}
\widehat G_\al
\ \text{if}\ \s=0, \mu\neq 0\ea$ \\
    \hline
    $2$ & $\ba{l}\lambda =0,\ \mu \neq \nu,
    \ea$ & $\ba[t]{l}e^{\nu
t}{\partial_ z
},\ e^{\mu t}\left( \sigma {\partial_ z }+\left( \mu -\nu \right)
R{\partial_ R}\right)\ea$&$\ba[t]{l}G_\al\ \text{if}\ \mu=\s=0,\ea$\\
&$\mu^2+\nu^2=1$&$\ba{l}{[}\texttt{AET 15}\ \texttt{if}\
\mu\nu=0{]}\ea$
  &$\ba[t]{l}
\widehat G_\al\ \text{if}\ \mu\neq
0, \s=0
\ea$
\\ \hline
$3$ & $\ba{l}\Delta=0,\\  \lambda=\varepsilon,\\
\mu+\nu=2\Omega\ea$ & $\ba{l}X_5=e^{\Omega t}\left( 2\varepsilon
R{\partial_ R}+( \nu -\mu ) {\partial_ z }\right),\\  2e^{\Omega
t}{\partial_ z }+tX_5\\{[}\texttt{AET 15}\  \texttt{if}\
 \nu+\mu=0,\\ \&\ \texttt{1, 17} \ \texttt{if}\ \mu=\nu=0{]}\ea $ &$\ba{l}G_\al\ \text{if}\  \mu=\nu= 0,\\
\widehat G_\al\ \text{if}\  \mu=\nu\neq 0
 \ea$  \\
     \hline
    $4$ & $\ba{l}\lambda\neq 0,\ \Delta=1\ea $ &
$\ba{l}e^{\omega _{+}t}\left( \lambda R{\partial_ R}+\left( \omega
_{+}-\mu \right)
{\partial_z }\right), \ea$ &$\ba{l}G_\al \ \text{if}\ \s=\mu= 0,
\ea$  \\
 & $\omega_\pm=\Omega\pm 1$ & $\ba{l}e^{\omega _{-}t}\left( \lambda R{\partial_ R}
 +\left( \omega _{-}-\mu
\right) {\partial_ z }\right)\\{[}\texttt{AET 15 } \texttt{if
}\mu\nu=\lambda\s, \\ \&\ 1 \texttt{ if }\mu=\s=0 {]}\ea$&
$\ba{l}\widehat G_\al \ \text{if}\ \s=0,\mu\neq 0\ea$
 \\
\hline $5$ & $\Delta=-1$ & $\ba{l} \exp (\Omega t)\left[2
\lambda\cos  tR{\p_R}\right.\\ \left.+\left( \left( \nu-\mu\right)
\cos  t- 2\sin t\right){\p_z} \right],\\\exp (\Omega t)\left[2
\lambda\sin tR{\p_R}\right.\\\left.+\left( \left( \nu-\mu\right)
\sin t+ 2\cos t\right){\p_z} \right] \ea $ & none   \\
\hline
\end{tabular}
\vspace{5mm}
\begin{center}
\textbf{Table 8.  Symmetries of equations (2) with $a=1$ and
non-linearities $f^1=\lambda v+\mu u \ln u$, $ f^2=\lambda \frac
{v^2}{u}+(\s u+\mu v)\ln u+\nu v,\ \lbd^2+\s^2\neq0$ }
\end{center}
\begin{tabular}{|l|l|l|l|}

\hline No&Conditions for&Symmetries and AET  Eq. (15)&Additional\\
& coefficients &&symmetries \\
\hline $1$&$\ba{l}\lbd=0,\\ \mu=\nu=\delta\ea$&$\ba{l}e^{\delta
t}u\p_v,\ e^{\delta t}(R\p_R+ \s t u\p_v)\\\texttt{{[AET 10}} \
\texttt{if}\ \mu=0{]}\ea$&$\ba{l}{\tilde\psi}_0\p_v,D+v\p_v,\\
\
\texttt{if}\ \mu=0\ea$\\
\hline
 $2$&$\ba{c}\lambda=0, \mu\neq \nu,\\\mu^2+\nu^2=1\ea$&$\ba{l} e^{\mu t}\lo (\mu-\nu)R{\p_R}+ \s u{\p_v}\ro
 ,\ e^{\nu
t}u{\p_v}\\{[}\texttt{AET 18 if}\ \mu\nu=0{]}
\ea$&$\ba{l}{\tilde\psi}_\nu{\p_v}\ \texttt{if}
\ \mu=0,\\\ea$\\
     \hline
    $3$ & $
\begin{array}{l}
\Delta =0,\\ \lbd=\varepsilon, \\
\mu +\nu =2\Omega
\end{array}
$ & $
\begin{array}{l}
X^4=e^{\Omega  t}\left( 2\varepsilon  R{\p_R}+(\nu -\mu
)u{\partial_v}\right) ,\\2e^{\Omega t}u{\partial_v}+tX^4\
 {[}\texttt{AET 18 if}\\ \mu=-\nu,\  \& \ \texttt{1, 19 } \texttt{if}\
 \mu=\nu=0{]}
 \ea
$ &$\begin{array}{l}D+u{\p_u}, G_a \\\text{if}\  \mu =\nu=0;\\\widehat{G}^a\ \text{if }\ \mu=\nu
\neq
0\end{array}$\\
    \hline
    $4$&$\ba{l}\lambda\neq 0,\\  \Delta=1,\ea $ &$
\ba{l}e^{\omega_+ t}\lo \lambda
R{\p_R}+(\omega_+-\mu)u{\p_v}\ro,\\e^{\omega_- t}\lo \lambda
R{\p_R}+(\omega_--\mu)u{\p_v}\ro\ea $&$\ba{l} G_a\ \text{if} \
\s=\mu=0,
\ea$\\
 &$\omega_\pm=\Omega\pm 1$&$\ba{l}{[}
\texttt{AET 18 if}\ \mu\nu=\lambda\s,\\ \&\ 1\  \texttt{if }
\mu=\s=0{]} \ea$&$\ba{l}\widehat G_\al \ \text{if}\ \s=0,\mu\neq 0
 \ea$\\
\hline
$5$ & $
\begin{array}{l}
\Delta =-1,
\end{array}
$ &$
\begin{array}{l}
e^{\Omega  t}[2\lambda \cos  tR{\p_R}\\+((\nu -\mu )\cos
 t
-2\sin t)u{\partial_v}], \\
e^{\Omega t}[2\lambda \sin tR{\p_R}\\ +((\nu -\mu )\sin
t +2\cos  t)u{\partial_v}]\ea$&none\\
\hline
\end{tabular}

\vspace{5mm}

Symmetries presented in Table 9 are related to equations (\ref{1.1})
with the unit diffusion matrix only.

\begin{center}{\bf
Table 9. Additional non-linearities with arbitrary parameters
and symmetries  for equation (2) with $a=1$}
\end{center}
\begin{tabular}{|l|l|l|l|}
    \hline No &$ \text{Nonlinear terms} $ &$ \text{Symmetries} $&
$\ba{l}\text{AET Eq.}\ (15) \ea$\\
\hline
$1.$&$\ba{l}f^1=\delta,\ f^2=u^\nu,\\
\delta=0\ \texttt{if }\nu=2\ea$&$\ba{l}D+u\p_u+(\nu+1)v\p_v,\\
{\tilde\psi}_0\p_v,\ \ (u-\delta t)\p_v,\\ \texttt{for any} \ \nu,
\delta, \  {\&}\\ u\p_u+\nu v\p_v \ \texttt{for}\ \delta=0 \\ \&\
\p_u+2tu\p_v\ \texttt{for}\
\nu=2\ea$&$\ba{l}3,9; {\&}\ 14,\\
\rho=\nu\omega \  \texttt{if}\\ \delta=0,\ \&\\21\ \texttt{if
}\nu=2\ea$\\\hline $2.$&$\ba{l}f^1=\va u,\
f^2=u^\nu,\\\nu\neq0,1\ea$&$\ba{l}u\p_u+\nu v\p_v,\
{\tilde\psi_0}\p_v, \ e^{-\va t}u\p_v\\\texttt{for any }\ \nu,\
{\&}\\e^{\va t}(u\p_v+\va \p_u) \ \texttt{for}\ \nu=2\ea$&$\ba{l}3;
 {\&
}\\ 14, \  \rho=\nu\omega \ea$\\\hline $3.$&$f^1=\eta v,\
f^2=-\frac{v^2}{u}$&
$\ba{l}D-v\p_v,\ u\p_u+v\p_v, \ G_\al,\\ K+(2-\eta)(t(\eta u\p_u\\
-(2+\eta)v\p_v)-u\p_v)\ea$&$\ba{c}\ \ \ 1\ea$\\\hline
$4.$&$\ba{l}f^1=\va
u^{\nu+1},\\ f^2=\va u^\nu v,\ \nu\neq0\ea$& $\ba{l}\nu D-u\p_u, \
v\p_v,\ u\p_v \\ \texttt{for any}\ \nu,\   {\&}\ (1+tu)\p_v\\
\texttt{for}\ \nu=1\ea$&
 $\ba{c}  6;{ \&}\\14, \omega=0\ea$\\\hline $5$. &$\ba{l} f^1=\delta u^{\nu
+1},\\f^2=u^\nu \left( \delta v+\mu u^\sigma \right) ,\\ \nu +\sigma
\neq 0,\ \mu\nu\neq 0 \ea$&$ \ba{l}\nu D-u{\partial_u}-\s
v{\partial_v},\  u{ {\partial_v}}
\ea $&$\ba{l} \ \ \ \ 6\ea$\\
\hline
$6$.&$
\begin{array}{l}
f^1=\delta \left( 2v- u^2\right) ^{\nu +\frac
12},\\
f^2=\delta u\left( 2v- u^2\right) ^{\nu +\frac
12} \\
+\mu \left( 2v- u^2\right) ^{\nu +1}
\end{array}
$&$\ba{l} 2\nu D
-u{\partial_u}-2v{\partial_v},\\
{\partial_u}+u{\partial_v}\ \texttt{for any}\ \nu,\mu
\\{\&}\  2 t(u\p_v+\p_u)+\p_v\\ \text{for}\
\nu=\frac12,\mu=0\ea $&$\ba{c}13;\ \&\ 14,\\\rho=2\om\\\text{if}\ \nu=0\ea$\\
\hline $7$. &$\ba{l} f^1=\delta e^{u},\\ f^2= ue^{u}\ea $&$\ba{l}
\delta(D-{{ {\partial_u})-}}u{\partial_v},\
{\tilde\psi}_0{\partial_v}\ea $&$\ \ \ \ 3$\\
\hline
$8$.&$
\begin{array}{l}
f^1=\eta e^{ 2v-u^2 }, \\
 f^2=\left( \eta u+\mu \right) e^{  2v-u^2
}
\end{array}
$&$\ba{l} 2 D-{\partial_v},\ {\partial_u}+u{{\partial_v}}\ea $&$\ \ \ 13$\\
\hline
$9$. &$
\begin{array}{l}
f^1=\delta u^{\nu +1}e^{ \frac{v}{u}},
\\f^2=e^{ \frac{v}{u}}(\delta
v
+\sigma u)u^\nu\ea$&$\ba{l}
D-u{{\partial_v},}\ \nu D-u{{\partial_u}}-v{{\partial_v}}\\\texttt{for any}
\ \nu,\ {\&}\
G_\alpha \ \text{ for}\  \nu =0\ea$&$\ba{c}\ \ \ 12\ea$ \\
\hline
 $10$. &$
\begin{array}{l}
f^1=e^{\nu z}R^\s (\delta u -\mu v),
\\f^2=e^{\nu z}R^\s (\delta v +\mu u)\ea$&$
\begin{array}{l}
\s D-u{{\partial_u}}-v{\partial_v},\\
\nu D-u{\partial_v}+v {\partial_u}\ \texttt{for any}\ \s,\\
{\&}\ G_\alpha \ \text{ for }\ \s =0\ea$&$\ba{c}\ \ \ 15\ea$\\
 \hline
 $11$.&$\ba{l}f^1=\va u\ln u,\\f^2=\va v\ln u\ea$&
 $\ba{l}e^{\va t}\lo u
{\partial_u}+  v {\partial_v}\ro,\ v\p_v,\ u\p_v,\\
\widehat G^\al \ea$&$\ba{c}6;\ \&\\ 14,\om=0\ea$\\\hline
  $12.$&$\ba{l}f^1=\delta,\\ f^2=\ln u\ea$&$\ba{l}D+u\p_u+v\p_v+t\p_v,\
  {\tilde\psi}_0\p_v\\\
 (u-\delta t)\p_v\ \texttt{for any}\ \delta,\\  {\&}\
 u\p_u+t\p_v\ \texttt{for}\ \delta=0\ea$&$\ba{l}3,9, \\ {\&}\ 6,\ 7\\
 \texttt{if} \ \delta=0\ea$\\\hline

\end{tabular}
\newpage

\begin{center}{\bf
Table 9. Continued}
\end{center}
\begin{tabular}{|l|l|l|l|}
\hline
$13$.&$\ba{l}
f^1=\va u^{\mu +1},\\f^2=\va u^\mu (v -\ln u),\\ \mu\neq 0\ea$&$\ba{l}
\mu D-u{{\partial_u}}-{{\partial_v}, }\ u{{\partial_v}}\\ \texttt{for any}
\ \mu\neq0,\\{\&}\ {{\partial_v}+t}u{{\partial_v}\ }\
\text{for}\ \mu =1
\end{array}
$&$\ba{c}\ \  \ 6\ea$ \\
\hline
$14$.&$
\begin{array}{l}
f^1=\delta \ln (2v-u^2), \\
f^2=\sigma (2v-u^2)^{1/2}\\+\delta u\ln (2v-u^2)
\end{array}
$&$
\begin{array}{l}
D+u{{\partial_u}}+2v{{\partial_v}}\\
+2\delta t\left( {{\partial_u}+}u{{\partial_v}}%
\right) , \\
{{\partial_u}+}u{{\partial_v}}
\end{array}
$&$\ \ \ 13$\\
\hline
$15$.&$ \ba{l}f^1= \va u^{\nu +1},\ \nu\neq-1,\\ f^2=
u^{\nu +1}\ln u \ea $&$ \ba{l}\nu D-( u{{\partial_u}}
+v{\partial_v}+\va u{{\partial_v}}%
) ,\\ {\tilde\psi}_0{\partial_v} \ea$&$\ \ \ \ 3$\\
\hline $16$.&$
\begin{array}{l}
f^1=\va u^{\nu +1},\   \nu\neq 1\\
f^2=\va u^\nu v \ +u\ln u
\end{array}
$&$
\begin{array}{l}
\nu D-u{\partial_u}-tu{{\partial_v}%
} \\
-(1-\nu )v{{\partial_v},}\ \ u{{\partial_v}}
\end{array}
$&$\ba{c}6; \&\ 5, \kappa=\va\\\texttt{if}\ \nu=0\ea $\\
\hline
 $17$.&$\begin{array}{l}
f^1= 2 v- u^2, \\
f^2=(\mu+  u) \left( 2 v- u^2\right)\\-\frac{\mu^2}{2} u,\ \mu\neq0
\end{array} $&$\ba{l}X^1=e^{\mu t}\lo 2 {\p_ u}+2 u{\p_
v}\right.\\\left.+\mu {\p_ v}\ro,\ \
tX^1+ e^{\mu t}{\p_ v}\ea$&$$ \\
\hline
$18$.&$\begin{array}{l}
f^1=  2v- u^2, \\
f^2=(\mu+ u) \left( 2 v- u^2\right)\\+\frac{1- \mu^2}{2}
u\end{array} $&$\ba{l}X^{\pm}=e^{(\mu\pm 1)t}\lo 2{\p_
u}\right.\\\left.+2 u{\p_ v}+(\mu\pm 1)
 {\p_v}\ro\ea$
&$\ba{c}13\ \texttt{if}\\\mu^2=1 \ea$\\
\hline $19$.&$\begin{array}{l}
f^1= 2v- u^2, \\
f^2=-\frac{1+\mu^2}{2} u\\+(\mu+ u) \left( 2v- u^2\right)
\end{array} $&$\ba{l}e^{\mu t}\lo 2\cos
t\lo{\p_u}+ u{\p_v}\ro\right.\\
\left.+(\mu \cos t-\sin t){\p_v}\ro,\\
e^{\mu t}\lo 2\sin t\lo{\p_u}+u{\p_v}\ro\right.\\
\left.+(\mu \sin t+\cos t){\p_v}\ro
\ea$&\\
\hline
\end{tabular}

\vspace{4mm}

We did not consider decoupled systems (\ref{1.2}) whose symmetries can be
 easily
found using the classification results of Dorodnitsyn \cite{dorod}
for a single diffusion equation.
We also did not
specify the case of linear systems (\ref{1.2}) when
\be \label{fin} f^1=\nu u+\mu v+\al,\ \ f^2=\s u+\lbd v +\omega.\ee
Equivalence transformations (\ref{x.2}) and 1--3 of (\ref{eqv}) make it
possible to specify
values of parameters in (\ref{fin}) by imposing the following conditions:
\be\label{fin1} \al=\omega=\lbd=0;\ \ \mu\s=0\ \texttt{or}\ \mu=\pm\s.\ee
Moreover, if the diffusion matrix $A$ is proportional to the unit matrix then equation
(\ref{1.2}), (\ref{fin1})
can be reduced to the case $f^1=f^2=0$.

The classification results present in the tables are valid also for equations
(\ref{1.2}) whose r.h.s. have the form
(\ref{fin}), (\ref{fin1}). However, to save a room we did not
indicated the standard additional
symmetries of linear equations, i.e.,
$U\p_u$ and $V\p_v$ where $U$ and $V$ satisfy the relations
\[U_t-\Delta U=f^1,\ \ V_t-a\Delta V=f^2.\]

The following last table completes the classification results for the case
of singular diffusion matrix.
\begin{center}{\bf Table 10\\
Additional non-linearities and symmetries for equations (2) with $a=0$}
\end{center}
\begin{tabular}{|c|c|c|c|}
  \hline
  No&Nonlinear terms&\begin{tabular}{l}
Argu- \\
ments\\ of \
$F^1$, $F^2$ \\
\end{tabular}&$\ba{l}\text{Symmetries}
\ \text{and AET}\ (15)\\\text{[in square brackets]}\ea$
\\
  \hline
 $1$ &$\ba{l}f^1=F^1+(\delta-\mu)u,\\
 f^2=F^2+\delta v\ea$&$v- u$&$\ba{l}e^{\delta t}\Psi_\mu(x)(\p_u+\p_v),
 \\
  \texttt{[AET}\ 11, \eta=1 \\ \texttt{if }
 \mu=\delta=0\texttt{]}\ea$\\
 \hline
 2&$\ba{l}f^1=e^{ u}F^1,\ f^2=e^{ u}F^2,\\
 F^2=1\ \texttt{if}\ \eta=0\ea$&$v-\eta u$&$ D-\p_u-\eta\p_v$\\
\hline
 3&$\ba{l}f^1=F^1,\\f^2=F^2+\eta v\ea$&$u$&$\ba{c}e^{\eta t}\Psi(x)\p_v\\
 {[\texttt{AET}\ 3\ \texttt{if}\ \eta=0]}\ea$\\
\hline
 4&$\ba{l}f^1=vF^1,\  f^2=F^2\ea$&$u$& $ D+v\p_v$\\
\hline
 5&$\ba{l}f^1=uF^1+\delta uv,\\f^2=F^2+\delta v\ea$&$v-\ln u$&$\ba{l}e^{\delta t}(u\p_u+\p_v)\\
 {[\texttt{AET}\ 5\ \texttt{if}\ \delta=0]}\ea$\\
\hline
 6&$\ba{l}f^1=u^{\nu+1}F^1,\\f^2=u^\nu F^2,\ \nu\neq0\ea$&$v-\ln u$&$\ba{l}
 \nu D-u\p_u-\p_v\ea$\\
\hline
  7&$\ba{l}f^1=F^1+\nu u,\ f^2=\eta\ea$&$v$&$\psi_\nu\p_u$\\
\hline
 8&$\ba{l}f^1=u^{\nu+1}F^1, \ f^2=0\ea$&$v$&$\ba{l}\nu D-u\p_u\
 \texttt{for any }\nu\\
 \&\ \psi_0\p_u\ \texttt{if}\ \nu=-1\\{[}\texttt{AET}\ 2,\ \&\ 14,\ \rho=0\\\texttt{if }\nu=0{]},\ea$\\
\hline
$9$&$\ba{l}f^1=v^{1+\lbd},\  f^2=\delta\ea$&&$\ba{l}D+v\p_v-
\lbd u\p_u,\
\psi_0\p_u\\{[}\texttt{AET 2}{]}\ea$\\\hline
$10$&$\ba{l}f^1=\delta e^u,\ f^2=e^u\ea$&&$\ba{l}D-\p_u,\
\Psi(x)\p_v\\{[}3;\ {\&} \  4\ \texttt{if}\ \delta=0{]}\ea$\\\hline
$11$&$\ba{l}f^1=\ln v,\ f^2=\va \ea$&&$\ba{l}D+u\p_u+v\p_v+
t\p_u,\\
\psi_0\p_u\ {[}\texttt{AET}\ 2{]}\ea$\\\hline
$12$&$\ba{l}f^1=\delta u^{\nu+1}v^{-1},\\
f^2= u^\nu\ea$&&$\ba{l}D+v\p_v,\ \nu D-u\p_u\\{[}\texttt{AET}\ 14,\
\nu\om=\rho\ea$
\\\hline
\end{tabular}

\vspace{3mm}

In  Table 10 $\phi$ is an arbitrary function of $v$.
In addition to the equivalence transformations indicated in the fourth
column, all the corresponding equations
  (\ref{1.2}) admit the AET $u\to u, v\to \varphi(v)$
where $\varphi$ is an arbitrary function of $v$.

\section{Discussion}

We have carried out the group classification of systems of coupled reaction-diffusion equations
(\ref{1.2}) with a diagonal diffusion matrix. The classification results are present in
Tables 2-10.
Moreover, symmetries of equation (\ref{1.3}) with a singular diffusion matrix and additional first
derivative terms are presented in Table 1.

The list of non-equivalent systems (\ref{1.2}) appears to be rather extended, especially for
 the unit diffusion matrix. Equations (\ref{1.2}) with invertible and non-unit diffusion
matrix $A$ have an essentially shorter
list of different symmetries. If the diffusion matrix is singular the number of
inequivalent equations appears to be the smallest one which is caused by the powerful equivalence
relations $u\to u, v\to \phi(v)$ where $\phi$ is an arbitrary function of $v$.

More exactly,  if matrix  matrix $A$ be of type 1, equation (\ref{x.0}),
then there exist 9 non-equivalent classes of equations (\ref{1.2})
defined up to arbitrary
functions and 19 classes of such equations defined up to
parameters. The related non-linearities and symmetries are
presented in Tables 4,5 and 10. The presented
extensions of the basic
symmetries (\ref{4.1}) have dimensions from 1 up to 3 and include neither Galilei
generators $G_\al$ nor
conformal generators $K$.

In addition, in Table 1 thirteen classes of equations with a singular diffusion matrix
and first derivative terms are presented.

For the case when matrix $A$ is of type 2, equation (\ref{x.0}), we
indicate in Tables 2-5 ten classes of equations defined up to
arbitrary functions and thirty five classes of equations defined up
to arbitrary or fixed parameters. Among them there are 7 Galilei
invariant systems, whose r.h.s terms are given in Table 2, Item 2;
Table 3, Items 2,3,4 and  Table 4, Items 1,2. In addition, there
exist two systems of type (\ref{1.2}) with a diagonal (but not unit)
diffusion matrix, which are invariant w.r.t. extended Galilei
algebra spanned on  $P^\mu, J^{\mu\nu}$ (\ref{4.1}) dilatation
operator
 and also generators $G_\al, K$ (\ref{2.6}). These equations correspond
 to the non-linearities present
  in Table 4, Item 1
 and have the following form
 $$\ba{l}\displaystyle u_t-\Delta u=\lambda u\lo{u}{v^{-a}}\ro^{\frac4{m(1-a)}},\\
 \displaystyle v_t-a\Delta v=\s v\lo{u}{v^{-a}}\ro^{\frac4{m(1-a)}}\ea$$
 and
 $$\ba{l} u_t-\Delta u=\lambda v^{\frac{4+m}4},\\
 \displaystyle v_t-\lo 1+\frac4m\ro\Delta v=0.\ea$$

 Finally, if the diffusion matrix is the unit one then we indicate
 98
 non-equivalent classes of
 equations, among them 21
 including arbitrary functions and 14 admitting Galilei generators.
 There is the only equation admitting
 extended Galilei algebra, the related non-linearities are given in
 Table 9, Item 3.

Consider examples of well known reaction diffusion equations which
appear to be particular subjects of our analysis.

The CGL equation (\ref{la}) with $\beta=0$ can be
rewritten as
\be\ba{l} u_t-\Delta_2u=u+(u^2+v^2)(\alpha v-u),\\
v_t-\Delta_2v=v-(u^2+v^2)(v+\alpha u).\label{GL}\ea\ee
where $u$ and $v$ are real and imaginary components of the
complex function $W$.

The r.h.s. of equations (\ref{GL}) has the form presented in Item 8 of Table
6 (with $\lambda=\nu=0$), and so in addition to basic symmetries $<\p_0,\ \p_1,\
\p_2, x_1\p_2-x_2\p_1>$ this system admits the symmetry
\be\label{LG} X=u\p_v-v\p_u.\ee

Using the anzatse
$$u=e^{i(x_1\cos\theta +x_2\sin\theta )}\tilde u,\ v=e^{i(x_1\cos\theta
 +x_2\sin\theta )}\tilde v$$
where $\tilde u$ and $\tilde v$ are functions of $t$ and $\omega$,
$\omega=x_1\sin\theta -x_2\cos\theta $, $\theta$ is a parameter, the system (\ref{GL})
can be reduced to the form
\be\label{GLR}\tilde u-\tilde u_{\omega\omega}=(\tilde u^2+\tilde v^2)(\alpha \tilde v-
\tilde u),\ \tilde v_t-\tilde v_{\omega\omega}=
(\tilde u^2+\tilde v^2)( \tilde v+\alpha \tilde u).\ee

Main symmetries of the reduced equation (\ref{GLR}) appear to be more extended
then of the CGL one. As is indicated in Item 11 of Table 9 equation
(\ref{GLR}) admits symmetry (\ref{LG}) and also the following one:
$$X_2=2D-u\p_u-v\p_v.$$

The primitive predator-prey system (\ref{prey})
 is a particular case of equation (\ref{1.1}) with
the non-linearities given in the first line of Table 2 where
however $-\mu=\nu=1, F^1=-F^2={\frac{u}{v}}$ . In addition to
the basic symmetries $<\p_t,\ \p_x>$ this
equation admits the (main) symmetry:
\[
X= D-u\partial_u -v
\partial_v.
\]

The $\lambda -\omega $ reaction-diffusion system
(\ref{d0})
and its symmetries was studied in
paper \cite{AR}. Our investigations confirm and complete the results of
\cite{AR}. First we recognize that this system is a
particular case of (\ref{1.1}) with non-linearities given in Item
11 of Table 6 with $\mu=\nu=0$. Hence it admits the five
dimensional Lie algebra generated by basic symmetries (\ref{4.1})
with $\mu, \nu=1,2$ and also the symmetry (\ref{LG}).
This is in accordance with results of paper \cite{AR} for
arbitrary functions $\lambda $ and $\omega $. Moreover, using
Table 9, Item $11$ we find that for the cases when
\begin{equation}
\lambda(R) ={\tilde\lambda} R^{\nu},\quad \omega =\sigma R^\nu
\label{d4}
\end{equation}
equation (\ref{d0}) admits additional symmetry with respect to
scaling transformations generated  by the operator:
\begin{equation}
X= \nu D-u\partial_u -v
\partial_v
 .  \label{zz5}
\end{equation}
The other extensions of the basic symmetries correspond to the
case when $\lambda(R)=\mu\ln(R), \omega(R)=\sigma\ln(R)$, the
related additional symmetries are given in Table 7, Items 1, 2, 5 where
$\nu=\lambda=0$.

Consider now the system (\ref{d1}). This system
 admits the equivalence
transformation 1 (\ref{eqv}) for $\rho =-\omega$. Choosing
$\rho=2k$ we transform equation  (\ref{d1}) to the form
(\ref{1.1}) where $a=-1$, $f^1=-2u^2v$ and $f^2=2v^2u$.
The symmetries corresponding to these non-linearities are given in
the first line of Table 4. For $m=2$ the symmetries are the most extended
and include two dilatations, two Galilei generators $G_\al$ $\al=1,\ 2$
and the conformal generator $K$. All these symmetries except $K$ are
valid for other numbers $m$ of independent
variables.

Symmetries of equations (\ref{d1}) for $m=1$ were
investigated in paper \cite{kra} whose results are in accordance with our
analysis.

The results of the present paper related to non-degenerated diffusion matrix can be compared
 with those of \cite{danil} and
\cite{chern1}, \cite{chern2}.

Paper \cite{danil} was apparently the first work were the problem
of group classification of equations (\ref{1.2}) with a diagonal diffusion
matrix was formulated and partially solved. However the
classification results presented in \cite{danil} include only a small part of ones
presented in Tables 2-10.

In papers \cite{chern1}, \cite{chern2} Lie symmetries of the same
equations and also of systems of diffusion equations with the unit
diffusion matrix were classified. The results present in those papers
 are much more advanced then the
pioneer Danilov ones, nevertheless they are also incomplete. In
particular, the cases presented above in Items 13-15 of
Table 6 and
 Items 1, 2 of Table 7  were not indicated in
\cite{chern2}.
Moreover, many of equations treated in
\cite{chern2} as non-equivalent ones, in fact are equivalent. For instance,
all versions 14, 15, 18 and 20 from Table 4
present in \cite{chern2} are equivalent one to another.

Notice that the results related to the group classification of
systems of nonlinear systems of reaction--diffusion equations are
presented in very compressed form and discussed in the survey
\cite{nik4}. The principally new points of the present paper in
comparison with \cite{nik4} are the following ones:
\begin{itemize}
\item In the present paper we give the completed list of admissible
equivalence transformations (\ref{eqv}) for all classified equations
(\ref{1.2}) whereas in \cite{nik4} only an {\it a priori} fixed
subclass of equivalence transformations was discussed.
\item For any particular system of equations (\ref{1.2})
whose nonlinear terms are given in the classification tables the
admissible equivalence transformations are  specified and presented
explicitly at the same tables while in \cite{nik4} the general
(incomplete) list of such transformations was presented only.
\item We use our knowledge of all admissible equivalence
transformation to reduce the number of non--equivalent versions of
systems (\ref{1.2}) to absolute minimum. In particular many of
quantities which define nonlinearities and are treated in
\cite{nik4} as arbitrary parameters are reduced to $\delta=0,\pm1,
\varepsilon =\pm1$ or $\eta=0,1$ and possible values of parameter
$a$ in the diffusion matrix $A$ are reduced to ones given by
equation (\ref{x.0}). .
\item Summarizing, in the present paper the problem of group
classification of systems of reaction--diffusion equations
(\ref{1.2}) is solved completely whereas all previous publications
\cite{danil}-\cite{chern2} and \cite{nik4} can be treated only as
steps to the complete solution.
\end{itemize}

Thus we present group classification of reaction-diffusion systems
with a diagonal diffusion matrix. Such systems with the square and
triangular diffusion matrix has been classified in paper \cite{nik1}
 and preprint \cite{nik3} respectively. The results of papers \cite{nik1}, \cite{nik3} and the
 present one consist in the completed group classification of systems of two coupled
 diffusion equations with the general diffusion matrix.

\appendix
\section{Appendix.
Algebras of main symmetries}

Following \cite{nik1} we first specify all non-equivalent terms
 \be \label{8.1}
N=C^{ab}u_b\p_{ u_a} + B^a \p_{ u_a}. \ee
where summation from 1 to 2 is imposed over the repeated indices and we again use the notations
$u_1=u, u_2=v$.

Let (\ref{8.1}) be a basis element of a one-dimensional invariance
algebra $\cal A$ then commutators of $N$ with  $P^0$ and $P^a$
should be equal to a linear combination
of $N$ and operators (\ref{4.1}). This condition presents
three the following possibilities \cite{nik1}:
\be \label{8.2} \ba{ll}1.& C^{ab}=\mu^{ab}, \quad
B^a=\mu^a, \\
 2. &C^{ab}=e^{\lbd t} \mu^{ab},
\quad B^a=e^{\lbd t} \mu^a,\\
 3.& C^{ab}=0, \quad
B^a= e^{\lbd t+\om \cdot x} \mu^a\ea \ee where $\mu^{ab}, \mu^a, \lbd
$, and $\om$ are constants.

Like in \cite{nik1} to classify all non-equivalent symmetries (\ref{8.2}) we use their
isomorphism
with $3
\times 3$ matrices of the following form \be \label{8.5} g= \left( \ba{ccc}
0   &  0       &  0\\
\mu^1 &  \mu^{11}  &  \mu^{12}\\
\mu^2 &  \mu^{21}  &  \mu^{12}  \ea \right). \ee

Equations (\ref{1.2}) admit equivalence transformations
(\ref{x.2}). The corresponding transformation for matrix
(\ref{8.5}) are \be \label{8.6} g \to g'=U gU^{-1},\ \ U=\left( \ba{ccc}
1   &  0       &  0\\
b^1 &  K^{11}  &  K^{12}\\
b^2 &  K^{21}  &  K^{22}  \ea \right)
 \ee
were $K^{ab}$ are the same parameters as in (\ref{x.2}), (\ref{x.}).

For the case of equation (\ref{1.2}) with $a\neq 1$ matrices
$\mu$ and $K$
in (\ref{8.5}), (\ref{8.6}) are diagonal, and up to equivalence
there exist there exist three matrices (\ref{8.5}), namely
\be \label{8.10}
g^1=\left( \ba{ccc}
0   &  0       &  0\\
0 &  1  &   0 \\
0 &  0  &  \lbd  \ea \right), \quad g^2=\left( \ba{ccc}
0   &  0       &  0\\
1 &  0  &   0 \\
0 &  0  &   1  \ea \right), \quad g^3=\left( \ba{ccc}
0   &  0       &  0\\
\lbd &  0  &   0 \\
1 &  0  &  0  \ea \right). \ee

In accordance with (\ref{4.2}), (\ref{8.1}),(\ref{8.2}) the related symmetry
operator can be represented in one of the following forms \be
\label{8.11}\ba{l} X^1_{(k)}=\mu D-2(g^k)_{bc}\tilde u_c \p_{
\tilde u_b},\ X^2_{(k)}=e^{\lbd t}(g^k)_{bc}\tilde u_c \p_{\tilde u_b},\\\\
X^3=e^{\lbd t+\om \cdot x}\left(
\p_{u_2}+\mu \p_{u_1}\right),\ k=1,2,3. \ea\ee Here
$(g^k)_{bc}$ are elements of matrices (\ref{8.10}), $b,c$ = 0, 1,
2, $\tilde u=$ column $(1, u_1, u_2)$.

Formulae (\ref{8.11}) and (\ref{8.10}) give the principal
description of one-dimension algebras $\cal A$ for equation
(\ref{1.2}) with $a\neq 1$.

To describe two-, three- and four-dimension algebras $\cal A$ we first
classify the corresponding algebras $A_{n,s}$ of matrices
$g$ (\ref{8.5}) where index $n$ indicates the
dimension of the algebra and $s$ is used mark different algebras of the same
dimension $n$.
Choosing
a basis element of $A_{2,s}$ in one of the forms given in (\ref{8.10})
we find that up to
equivalence transformations (\ref{8.5}) there exist six two dimension
algebras with basis elements
$<e_1, e_2>$: \be \label{8.13}\ba{l} A_{2,1}:\ e_1=  g^1_{(0)},
\ e_2=g^4;\ \ \ \
A_{2,2}:\ e_1=g^1_{(0)}, \ e_2=g^3_{(0)};\\ A_{2,3}:\ e_1=g^5,\ e_2=
g^3_{(0)},\ea\ee \be \label{8.14}\ba{l} A_{2,4}:\ e_1= g^1, \ e_2=g^5;\ \ \ \
 A_{2,5}:\ e_1= g^1_{(1)},\ e_2= g^3;\\ A_{2,6}:\ e_1=g^2,\ e_3=  g^3_{(0)} \ea
\ee where $ g^1_{(0)}=g^1|_{\lbd=0}$, $g^1_{(1)}=g^1|_{\lbd=1}$,
$g^3_{(0)}=g^3|_{\lbd=0}$, and \be \label{8.15} g^4=\left(
\ba{ccc}
0   &  0       &  0\\
0 &  0  &   0 \\
0 &  0  &   1 \ea \right), \quad g^5=\left( \ba{ccc}
0   &  0       &  0\\
1 &  0  &   0 \\
0 &  0  &   0  \ea \right). \ee

Algebras (\ref{8.13}) are Abelian while algebras (\ref{8.14}) are
characterized by the following commutation relations: \be\label{L}
[e_1, e_2]= e_2.\ee

Using (\ref{8.13}), (\ref{8.14}) and applying arguments analogous
to those which follow equations (\ref{5.2}) we find pairs
of operators (\ref{4.2}) forming Lie algebras. Denoting
\[
\hat e_\al=(e_\al)_{ab} {\tilde u}_b \frac{\p}{\p \tilde u_a}, \quad
\al=1,2
\]
we represent them as follows: \be \label{8.17}\ba{l} < \mu D+ \hat
e_1+\nu  t \hat e_2, \hat e_2>,\ \ \ < \mu D+ \hat e_2+\nu  t \hat
e_1, \hat e_1>,\\<\mu D-\hat e_1,
         \nu D-\hat e_2>,\ \ \   <F^1 \hat e_1+ G^1 \hat e_2,\ F^2\hat e_1+G^2
 \hat e_2>\ea \ee for $e_1, e_2$ belonging to algebras (\ref{8.13}) and
\be \label{8.18} <\mu D- \hat e_1, \hat e_2>, \ \ \
 < \mu D+ \hat e_1+\nu  t \hat e_2, \hat e_2>\ee for $e_1, e_2$
belonging to algebra (\ref{8.14}).

Here $\{F^1, G^1\}$ and $\{F^2,G^2\}$ are fundamental solutions of
the following system \be \label{8.20}  F_t=\lbd F+\nu G, \quad
G_t = \sigma F+\gm G \ee with arbitrary parameters $\lbd, \nu, \s,
\gamma$.

The list (\ref{8.17})-(\ref{8.18}) does not includes two-dimension algebras whose basis is
$< F\hat e_\al,\ G\hat e_\al>$ (with $F, G$ satisfying (\ref{8.20})) or
$<\mu D+\lambda e^{\nu t +\om \cdot x}\hat e_\al, e^{\nu t +\om \cdot x}
\hat e_\al>$
which are incompatible with classifying equations (\ref{2.7}). In the following we
ignore all algebras $\cal A$ which include
such subalgebras.

Up to equivalence there exist three realizations of
three-dimension algebras of matrices (\ref{8.10}),
(\ref{8.15}): \be \label{8.21}\ba{l} A_{3,1}:\ \ e_1=g^1_{(0)}, \
e_2=g^4,\ e_3=g^3_{(0)}, \\  A_{3,2}:\ \ e_1=g^5,\ e_2= g^4,\ e_3=
g^3_{(0)}, \ea\ee \be\label{8.22}\ba{l} A_{3,3}:\ \ e_1=g^1_{(1)}, \
e_2=g^5, \ e_3=g^3_{(0)}.\ea \ee

Non-zero commutators for matrices (\ref{8.21}) and (\ref{8.22})
are $[e_2,e_3]=e_3$ and $[e_1,e_\al]=e_\al (\al=2,3)$ respectively. The
algebras of operators (\ref{4.2}) corresponding to realizations
(\ref{8.21}) and (\ref{8.22}) are of the following general forms:
\be\label{8.23}\ba{l}<\mu D-\hat e_1,\  \nu D-\hat e_2, \hat
e_3>, \
  <\hat e_1,  \ D+\hat e_2+\mu t \hat e_3,\ \hat e_3>\ea\ee
and \be\label{8.24}\ba{l}<\mu D-\hat e_1,\ \hat e_2,\  \hat e_3>,
\
<D+\hat e_1+\nu t\hat e_2,\  \hat e_2,\ \hat e_3>,\\
<D+\hat e_1+ \nu t\hat e_3,\  \hat e_3,\ \hat e_2>,\ <\hat e_1,\
\ F^1 \hat e_2+G^1 \hat e_3,\ F^2\hat e_2+G^2 \hat e_3>\ea\ee
correspondingly.

In addition, we have the only four-dimension algebra
\begin{equation}\hat
A_{4,1}:\ \ e_1=g^1_{(0)},\ e_2=g^5,\ e_3=g^3_{(0)},\
e_4= g^4\label{1111}\end{equation} which generates the following
algebras of operators (\ref{4.2}):
\begin{equation}\label{8.25} \ba{l}<\mu D-\hat e_1,\ \nu D-\hat
e_3,\ \hat e_2, \ \hat e_4>,\ \
<\hat e_1, \ D+\hat e_3+\nu t \hat e_4, \ \hat e_2,\ \hat e_4>,\\
<D+\hat e_1+\nu t \hat e_2,\ \hat e_2,\ \hat e_3,\ \hat e_4>.\ea
\end{equation}

Finally, it is necessary to take into account the special type of
(m+2) -dimensional algebras $\cal A$
 generated by two-dimension algebras
(\ref{8.13}), namely, algebras whose basis elements have the following general form:
$<\mu D+\hat e_1+(\al t+\lambda^{\s\rho} x_\s x_\rho)\hat e_2, x_\nu \hat e_2, \hat e_2>$
where $\nu,\s,\rho$ run from 1 to $m$. The related classifying equations generated by
all symmetries $ x_1\hat e_2, x_2\hat e_2, \cdots,  x_m\hat e_2$ and  $\hat e_2$ coincides
and we have the same number of constrains for $f^1,f^2$ as in the case of two-dimension
algebras $\cal A$.

The case $a=1$ appears to be much more
complicated. The related matrices $g$ are of general form
(\ref{8.5}) and defined up to the general equivalence
transformation (\ref{8.6}) with arbitrary $K^{ab}$. Namely
there are  seven non-equivalent matrices (\ref{8.5}), including
$g^1, g^2$ (\ref{8.10}), $g^5$ (\ref{8.15})
 and also the following matrices
\be\label{8.28}\ba{l}
g^6=\lo\ba{ccc}0&0&0\\0&\mu&-1\\0&1&\mu\ea\ro, \ \ g^7=\lo\ba{ccc}0&0&0\\0&1&0\\0&1&1\ea\ro, \\
\\
g^8=\lo\ba{ccc}0&0&0\\0&0&0\\0&1&0\ea\ro, \ \
g^9=\lo\ba{ccc}0&0&0\\1&0&0\\0&1&0\ea\ro.\ea\ee
 In addition, we have
fifteen two-dimension algebras of matrices (\ref{8.5}),
\be\label{8.42}\ba{l}
A_{2,1}=<g^1_{(0)},\ g^4>, \ A_{2,2}=<g^1_{(0)},\ g^3_{(0)}>,\
A_{2,3}=<g^3_{(0)},\ g^5>, \\ A_{2,7}=<g^7,\ g^8>, \
A_{2,8}=<g^3_{(0)},\ g^{8}>, \ A_{2,9}=<g^3_{(0)},\ g^{9}>,
\\  A_{2,10}=< g^1_{(1)},\ g^6>,\ea\ee
\be\label{8.43}\ba{l}
A_{2,4}=< g^1,\ g^5>, \ A_{2,5}=< g^1_{(1)},\ g^3>,\  A_{2,6}=<g^2,\
g^3_{(0)}>, \\ A_{2,11}=<\frac1{\lambda-1}g^1|_{\lambda\neq 1},\ g^8>, \
A_{2,12}=< -g^{10}, \ g^{8}>, \ A_{2,13}=< {g}^{1}_{(2)},\ g^9>, \\
  A_{2,14}=< g^7,\ g^3_{(0)}>\ea\ee
where
$$g^{10}=\lo\ba{lll}0&0&0\\0&1&0\\1&0&0\ea\ro, \
g^1_{(2)}=g^1|_{\lambda=2}=\lo\ba{lll}0&0&0\\0&1&0\\0&0&2\ea\ro.$$

Algebras (\ref{8.42}) are Abelian whereas  algebras (\ref{8.43}) are
characterized by commutation relations (\ref{L}). The corresponding
algebras $\cal A$
are given by equation (\ref{8.17}) and (\ref{8.18}) respectively.

Three-dimension algebras $A_{3,s}$ are the algebras $ A_{3,1}- A_{3,3}$
given by relations
(\ref{8.21}), (\ref{8.22}) where matrices $g^1$ and $g^3$ are of general
form (\ref{8.10}) with arbitrary $\lambda$ (i.e., $g^1_{(0)}\to g^1$, etc.)
and also algebras $A_{3,4}-A_{3,11}$ given below:
\[ \ba{l}A_{3,4}: \ \ e_1=g^8, \ e_2=g^{1}_{(1)}, \ e_3=g^3_{(0)},\\
A_{3,5}:\ \
e_1=g^1,\ e_2=g^8,\ e_3=g^3_{(0)},\\
A_{3,6}: \  e_1= g^1_{(0)},\ e_2=g^8,\ e_3= g^4,\\
A_{3,7}: \ \ e_1=g^4,\ e_2=g^8,\ e_3=g^3_{(0)},\\
A_{3,8}:\ \ e_1=g^5,\ e_2=g^6,\ e_3=g^3_{(0)},\\
A_{3,9}: \ \ e_1=g^3_{(0)}, \ e_2=g^8, \ e_3=g^9,\\
A_{3,10}: \  e_1=g^2,\ e_2=g^8,\ e_3=g^3_{(0)},\\
A_{3,11}: \ \ e_1=g^3_{(0)}, \ e_2=g^5, \ e_3=g^7
.\ea\]

Algebras $A_{3,4}- A_{3,6}$ and  $A_{3,7}$ are
isomorphic to $A_{3,1}$ and $A_{3,3}$ respectively.
The related algebras $\cal A$ are given by equations (\ref{8.23}) and
(\ref{8.24}) correspondingly.

Algebra $A_{3,8}$ is isomorphic to $A_{3,3}$ and so generates algebra (\ref{8.24}).

Algebras $A_{3,9}$ and $A_{3,10}$ are characterized by the following commutation relations
\be\label{8.37}[e_2,e_3]=e_1\ee
(the remaining commutators
are equal to zero); non-zero commutators for basis elements of $A_{3,11}$ are given below:
\be\label{8.38}
[e_1,e_2]=e_2,\ \ [e_1,e_3]=e_2+e_3.\ee
Using (\ref{8.37}) and (\ref{8.38}) we come to the following related three-dimension
algebras $\cal A$ generated by $A_{3,9}$ and $A_{3,10}$:
\be\label{8.39}\ba{l}
<\mu D-2\hat e_2,\ \nu D-2\hat e_3,\ e_1>, \
<e_1,\ D+2e_\al+2\nu t e_1,\ e_{\al'}>,\\
<e^{\nu t}e_1,\ e^{\nu t}e_\al,\
e_{\al'}>,\ \ \ \al,\al'=2,3,\ \al'\neq \al  \ea \ee
and algebras (\ref{8.40}) generated by $A_{3,11}$:
\be\label{8.40}<\mu D-2e_1, \ e_2,\ e_3>,\ <e_1, \ e^{\nu
t}e_2,\ \ e^{\nu t}e_3>.
\ee

Finally, four-dimension algebras of matrices (\ref{8.6}) are $A_{4,1}$
 given by equations (\ref{1111}) and also
$A_{4,2}$--$A_{4,5}$ given below:
$$\ba{l} A_{4,2}:\ \ e_1=g^{1'},\ e_2=g^6,\ e_3=g^3_{(0)},\
e_4=g^5;\\
A_{4,3}: \ \ e_1=g^3_{(0)}, \ e_2=g^5, \ e_3=g^{1'},\ e_4=g^8;\\
A_{4,4}: \ e_1=g^1,\ e_2=g^4,\ e_3= g^8,\ e_4=g^3;
\\A_{4,5}: \ e_1=g^4,\ e_2=g^8,\ e_3= g^5,\ e_4=g^3.\ea$$

We do not present the related algebras $\cal A$ because all possible non-linearities
$f^1$ and $f^2$ will be fixed asking for invariance of equation (\ref{1.2}) which respect to transformations
generated by three-dimensional algebras.


\begin{thebibliography}{99}
\bibitem{nik1} A. G. Nikitin. {\it Group classification of systems of non-linear reaction-diffusion equations
with general diffusion matrix. I. Generalised Ginzburg-Landau equations},
J. Math. Anal. and Appl.(JMAA) {\bf 324}, 615-628, 2006; ArXiv math-ph/0411027,
 2004.

 \bibitem{turing} A. M. Turing, {\it The chemical basis of morphogenesis},
 Phil. Trans. Roy. Soc. London B., {\bf 237} 37-72, 1952.
 \bibitem{nik3} A. G. Nikitin, {\it Group classification of systems of non-linear reaction-diffusion equations
with general diffusion matrix. III. Triangular diffusion matrix}.
ArXiv math-ph/0411028, 2004
\bibitem{danil} Yu. A. Danilov, {\it Group analysis of the Turing systems
and of its analogues.} Preprint of Kurchatov Institute for Atomic Energy
IAE-3287/1, 1980.
\bibitem{nikwil2}A. G. Nikitin and R. Wiltshire, {\it Symmetries of Systems of
     Nonlinear Reaction-Diffusion Equations},
in:  Symmetries in Nonlinear Mathematical Physics, Proc. of
the Third Int. Conf. , Kiev, July 12-18, 1999, Ed. A.M.
Samoilenko ( Inst. of Mathematics of Nat. Acad. Sci. of Ukraine,
Kiev, pp. 47-59, 2000).
\bibitem{nikwil1}A. G. Nikitin and R. Wiltshire, {\it Systems of Reaction
     Diffusion Equations and their symmetry properties}, J. Math. Phys. {\bf
42} 1667-1688, 2001.
\bibitem{kniazeva}I. V. Knyazeva and M. D.Popov, Group classification of diffusion equations,
Preprint 6 of Keldysh Institute of
Applied Mathematics, USSR Academy of Sciences, Moscow, 1986.
See also:
 CRC Handbook of Lie Group Analysis of Differential Equations, Vol. 1, (Ed. N.Ibragimov), CRC
Press, 1994, p.171-176.
\bibitem{chern1} R. M. Cherniha and J. R. King, {\it Lie
symmetries of nonlinear multidimensional reaction–-diffusion
systems: I }, J. Phys. A {\bf
33}, 267-282, 2000;\\
R. M. Cherniha and J. R. King, {\it Lie symmetries of nonlinear
multidimensional reaction-–diffusion systems: I. Addendum },  J.
Phys. A{\bf 33}, 7839-7841, 2000.
\bibitem{chern2}R. M. Cherniha and J. R. King,
{\it Lie symmetries of nonlinear multidimensional
reaction-–diffusion systems: II }, J. Phys. A {\bf 36}, 405-425,
2002.
\bibitem{martina}I. Martina, O. K. Pashaev and G. Soliani, {\it Bright solitons as black holes}, Phys. Rev.
D {\bf 58}, 084025, 1998.
\bibitem{kra}R. A. Kraenkel and M. Senthilvelan,
{\it Symmetry analysis of an integrable reaction-
diffusion equation, } Ch. Sol. Fract.
{\bf 12}, 463-474, 2001.
\bibitem{murry} J. D. Murray. Mathematical Biology. Springer, 1991.
\bibitem{Kop} N. Kopell and L. N. Howard, {\it Plane wave solutions to reaction-diffusion equations}, Studies in Appl. Math.
{\bf 52}, 291, 1973.
\bibitem{green} J. M. Greenberg, {\it Spiral waves for
$\lambda-\omega$ systems}, Adv. Appl. Math. {\bf 2}, 450,
1981.
\bibitem{AR}  J. F. R. Archilla et al, {\it Lie symmetries and multiple
solutions in $\lambda-\omega$  reaction-diffusion systems },
J. Phys. A \textbf{39,} 185
(1997).
\bibitem{fitz}R. Fitzhugh, {\it Impulses and Physiological States in
Models of Nerve Membrane}, Biophys. J. {\bf 1}, 445, 1961
\bibitem{rinzel}J. Rinzel and J. B. Keller, {\it Travelling wave solutions
 of a nerve conduction equation}, Biophys. J. {\bf 13}, 1313, 1973.
\bibitem{dorod} V. A. Dorodnitsyn, {\it Group properties and invariant solutions of a linear heat equation
with a  source.} Preprint of Keldysh Institute of Applied Mathematics,
Academy of Sci. of U.S.S.R
{\bf 57}, 1979.
\bibitem{nik4} A. G. Nikitin, {\it Group classification of systems of
non-linear reaction--diffusion equations}, Ukrainian Mathematical
Bulletin {\bf 2}, 153-204, 2005.
\end{thebibliography}
\end{document}